%% file: main.tex
\documentclass[a4paper,fleqn,usenatbib]{mnras}
\usepackage{booktabs,caption}
\usepackage{graphicx,epsfig,epsf,graphics,epstopdf}
\usepackage[flushleft]{threeparttable}
\usepackage{indentfirst}
\usepackage{epsfig}
\title[Tailed Radio Galaxies from TGSS]{Tailed Radio Galaxies from the TIFR GMRT Sky Survey}
\author[Bhukta et al.]{
Netai Bhukta$^{1}$,
Sushanta K. Mondal$^{1}$,
Sabyasachi Pal$^{2}$\thanks{E-mail: sabya.pal@gmail.com}
\\
$^{1}$Department of Physics, Sidho Kanho Birsha University, Ranchi Road, Purulia 723104, India\\
$^{2}$Indian Centre for Space Physics, 43-Chalantika, Garia Station Road, Kolkata 700084, India\\
}
\date{Accepted XXX. Received YYY; in original form ZZZ}
\pubyear{2016}
\begin{document}
\label{firstpage}
\pagerange{\pageref{firstpage}--\pageref{lastpage}}
\maketitle
\begin{abstract}
We present a list of candidate tailed radio galaxies using the TIFR GMRT Sky Survey Alternative Data Release 1 (TGSS ADR1) at 150 MHz. We have visually examined 5336 image fields and found 264 candidates for the tailed galaxy. Tailed radio galaxies are classified as Wide Angle Tailed (WAT) galaxies and Narrow-Angle Tailed (NAT) galaxies, based on the angle between the two jets of the galaxy. We found a sample of tailed radio galaxies which includes 203 `WAT' and 61 `NAT' type sources. These newly identified tailed sources are significant additions to the list of known tailed radio galaxies. The source morphology and luminosity features of the different candidate galaxies and their optical identifications are presented in the paper. Other radio properties and general features of the sources are also discussed.
\end{abstract}

\section{Introduction}
\label{sec:intro}
Tailed radio galaxies are galaxies having a pair of radio ``tails" extended far from the optical galaxy. The radio ``tails'' are nothing but the bending of both of the radio jets in the same direction. Initially, they are well-collimated on a kiloparsec scale and suddenly flare into diffuse plumes, which may be significantly bent \citep{Mi19, Pat19}. The spectral index of tailed radio galaxies becomes steeper moving outwards towards the tail from the core and varies from region to region \citep{Br81, Ha03, Pat19}. Most tailed radio galaxies are Fanaroff–Riley I radio galaxies \citep{Fr74}, where the surface brightness is relatively higher towards the end of the jets than the regions close to the core \citep{Te17, Mi19}. Tailed radio galaxies are distinguished by their radio morphology, which includes bright hotspots (called ``warmspots") that are closer to their radio core than FR IIs \citep{Mi19}. Tailed sources are classified depending on the angle between the radio tails and the core of the galaxy. The narrow-angle tail (NAT) radio sources are featured by tails bent in a narrow `V' or `L' shape where the angle between two tails is less than 90 degrees. The jet bending in the case of Wide-angle tail (WAT) radio sources is such that the WATs exhibit wide `C' type morphologies and the angle between the two components is greater than $90$ degrees but less than $180$ degrees. These `WAT' and `NAT' morphologies were first defined in \citet{Ow76}. The structures of NAT sources may be affected by the projection effect. The luminosity of the WAT sources comes in between the classical double and the NAT sources \citep{Od93}.

Tailed sources are generally found in the dynamical, active, and X-ray intense regions of the rich clusters \citep{Bur81, Od85}. It is considered that the ram pressure associated with the dynamical interaction of the host galaxy with the dense intercluster medium (ICM; assumed to be at rest) causes the radio jets to bend in the reverse direction of motion, away from the cluster centre \citep{Beg79, Val81, Baa85}. The degree of bending depends on the velocity of the host galaxies \citep{Sr20}. The NAT sources are found towards the edge of the associated cluster, while the WAT sources are generally found close to the gravitational centre of the clusters \citep{Qui82}. The ram pressure explanation of jet bending does not hold for the WAT sources as they move very slowly compared to the NATs \citep{Bur82a}. The WAT morphology is believed to cause by strong intercluster wind \citep{Bu86}, in some literature it is attributed to the electromagnetic interaction of the non-neutral jet with the ICM magnetic field \citep{Elk84}.

Tailed radio sources can be used to trace galaxy clusters and high redshift systems \citep{Bla00, Bla03, Smo07}; to investigate the inner cluster environment \citep{Dou11, Win11, Bla01}; to study the cluster magnetic field \citep{Fer99}; to study the interaction of the jet with the intercluster medium and to study the evolution of galaxies and galactic dynamics \citep{Pin92}.
The tailed radio galaxies are normally representatives of \citet{Fr74} class I sources, but in terms of luminosities, one can place them close to the FR II transition \citep{Mi72, Bl98}. However, on a scale of arcseconds, these tailed galaxies normally have U-shaped symmetry constructed by a pair of bent jets which originate from the nucleus and meet with the more elongated tails after adequate bending \citep{Ow79, Od86}. So far, most of the tailed galaxies have been discovered entirely in galaxy clusters, with the majority of them being rich galaxy clusters \citep{Ma09}. 
At least a few factors influence the bending of tailed radio galaxies.
(1) Most of the tailed radio galaxy search activities focus on galaxies in clusters \citep{Od85} and the bending of radio tails is due to the motion of the host galaxy through the inter-cluster medium.
(2) The hypotheses that host galaxy orbits are either radial, circular, or isotropic \citep{Jo79} are tested using the initial ejection angle of the jet with respect to the motion of the galaxy and the jet flow velocity \citep{Baa85}. 
(3) The differences in the ejection of jets with respect to the direction of motion of the host galaxy in the inter-cluster medium and environments, as well as projection effects, are responsible for the asymmetries found in the radio jets in some tailed radio galaxies \citep{Se17}.
(4) The probability of detection of tailed radio sources decreases in the clusters with less number of galaxies \citep{St77, Ad80}.
(5) Various types of bending may be caused by precessing radio jets.
Due to projection effects, the distinction between WATs and NATs is somewhat arbitrary. Many of the WAT sources look like NAT sources simply due to projection effects. Transonic or supersonic relative motions drive the jets of tailed radio galaxies. Even with high-resolution (arcsec scale) radio observations, a few NAT sources have been discovered to have a remarkable narrow extended structure, with the two radio tails not being resolved (e.g. IC 310; \citet{Fe88}).

WATs display bending of two jets less than NATs, and the bending model of dynamical ram pressure can not completely explain the bending of WATs morphology \citep{Bur81, Bur82b, Elk84}. WAT sources appear to be associated with large size D or cD galaxies approaching rest at cluster centres (e.g., 3C 465) and therefore should not have the necessary velocity to produce the ram pressure required for the observed bends in their radio jets. So, an alternative fundamental mechanism is required to interpret the noticed bending of jets of WATs. An electromagnetic force that results from the interaction of a jet carrying a net electrical current with the magnetic field of the inter-cluster medium may be responsible for the bending of jets in WATs \citep{Elk84}. An ordered magnetic field is required to generate the symmetric shape of WATs. Collisions with dense clouds in the ICM, on the other hand, may also deflect jets. In some WATs, this method may be responsible \citep{Bu86} but it is difficult to reproduce the large-scale symmetric structure of WATs by only this method.

The tailed radio galaxies are normally representatives of \citet{Fr74} class I sources, but in terms of luminosities, one can place them close to the FR II transition \citep{Mi72, Bl98}. However, on a scale of arcseconds, these tailed galaxies normally have U-shaped symmetry constructed by a pair of bent jets which originate from the nucleus and meet with the more elongated tails after adequate bending \citep{Ow79, Od86}. So far, most of the tailed galaxies have been discovered entirely in clusters, with the majority of them being rich clusters \citep{Ma09}.

\citet{Od85} has identified 57 tailed sources in the directions of different clusters using the A and B configurations of the VLA at $20$ cm. This sample of 57 sources includes $41$ NATs, $9$ WATs and $7$ sources with complex morphologies. \citet{Odn90} identified 11 WATs in $20$ cm using VLA in A and C configuration.  Detail study of big WAT 1919+479 is presented in \citet{Pi98}. Six other WATs are found in the ATLAS field at 1.4 GHz using ATCA \citep{Mao2010}. NGC 1265 is a well studied NAT source \citep{Xu99}. Around 1600 sources are identified as possible `tailed' candidates using a pattern recognition algorithm \citep{Pro16} using the NRAO VLA Sky Survey (NVSS; \citet{Con98}) at 1.4 GHz. 
In this paper, we present $264$ tailed radio sources of which $198$ sources are located in the northern sky (above 0 degrees). We have classified 203 sources as `WAT' type and 61 sources as `NAT' type based on the angle made by the two bent lobes. Most of these sources have been observed before and catalogued in different radio surveys, mostly in the NVSS survey and in the Sydney University Molonglo Sky Survey (SUMSS; \citet{Mau03}) at 843 MHz, but have not been reported as tailed sources. We found that only about half of the sources are associated with a known galaxy cluster. The optical galaxy hosting the radio sources is located in the redshift range of $0.01$ to $0.68$ and the total flux at 150 MHz ranges from 0.1 Jy to as large as 20.1 Jy. Redshifts are found for 165 WATS out of 203 detected optical/IR counterparts (75 per cent). For NATs, redshifts are found for 49 galaxies out of 61 detected optical/IR counterparts (76 per cent). Out of 117 identified redshifts of tailed radio galaxies, fifty are identified spectroscopically.
We arrange the paper in the following ways: In section \ref{sec:data}, we present the method of the identification of sources. We describe the counterpart identification in section \ref{subsection:counterpart}. In the next section (section \ref{sec:result}), we describe the different radio properties of the sources. In section \ref{sec:disc}, we discuss the general features and overall properties of sources. We summarise the study in the final section. We assumed the flat $\Lambda CDM$ cosmology with $H_0=67.8 \, \rm km \, s^{-1} \,Mpc^{-1}$, $\Omega_{\rm M}=0.308$, and $\Omega_\Lambda=0.692$ \citep{Ag18}. 

\section{Identication of tailed-radio galaxies}
\label{sec:data}
\subsection{The TGSS alternative data release one}
The tailed sources are found from the manual inspection of a large number of high-resolution images generated by the TIFR GMRT Sky Survey Alternative Data Release 1 (TGSS ADR 1; \citet{Int17}). The continuum survey at 150 MHz using the Giant Metrewave Radio Telescope (GMRT; \citet{Swa91}) covers a declination range from $-55$ to +$90$ degrees. The purpose of the survey was to provide a high-resolution and high sensitivity map of the 150 MHz sky. The median noise of the survey is 3.5 mJy beam$^{-1}$ and the resolution is $25\arcsec \times 25\arcsec$ north of 19$\degr$ DEC and $25\arcsec \times 25\arcsec / \cos(\textrm{DEC}-19\degr)$ south of 19$\degr$. In total, over 2000 hours of observation time were used over about 200 observing sessions. Earlier, four tailed galaxies were serendipitously discovered with GMRT at 610 and 327 MHz \citep{Gia09}. Detection of a WAT (J0037+18) with an interacting host galaxy as optical counterpart \citep{Pat19} is reported in the same frequency bands using GMRT.

\begin{figure*}
\includegraphics[width=8.4cm,angle=0,origin=c]{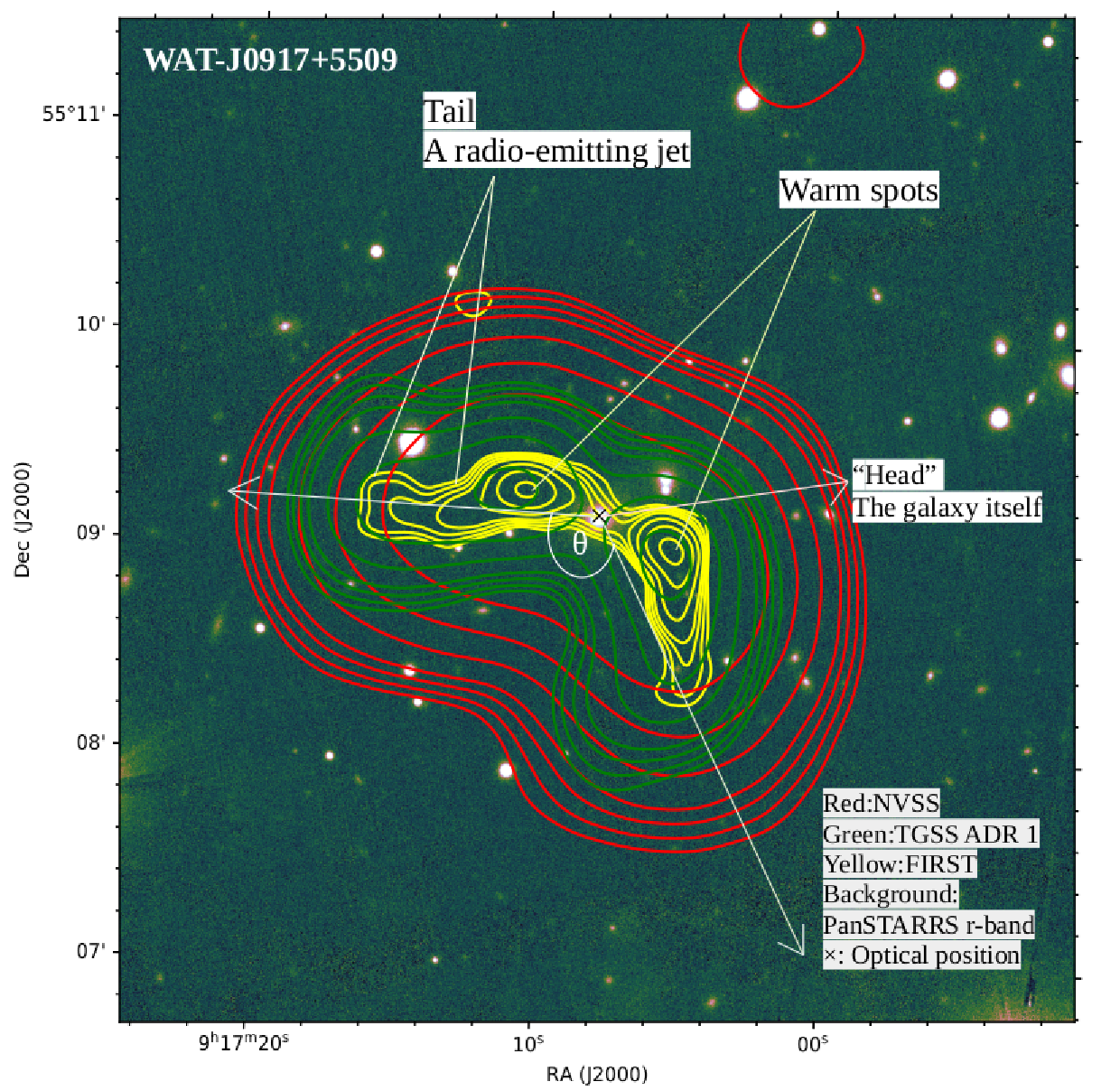}
\includegraphics[width=8.9cm,angle=0,origin=c]{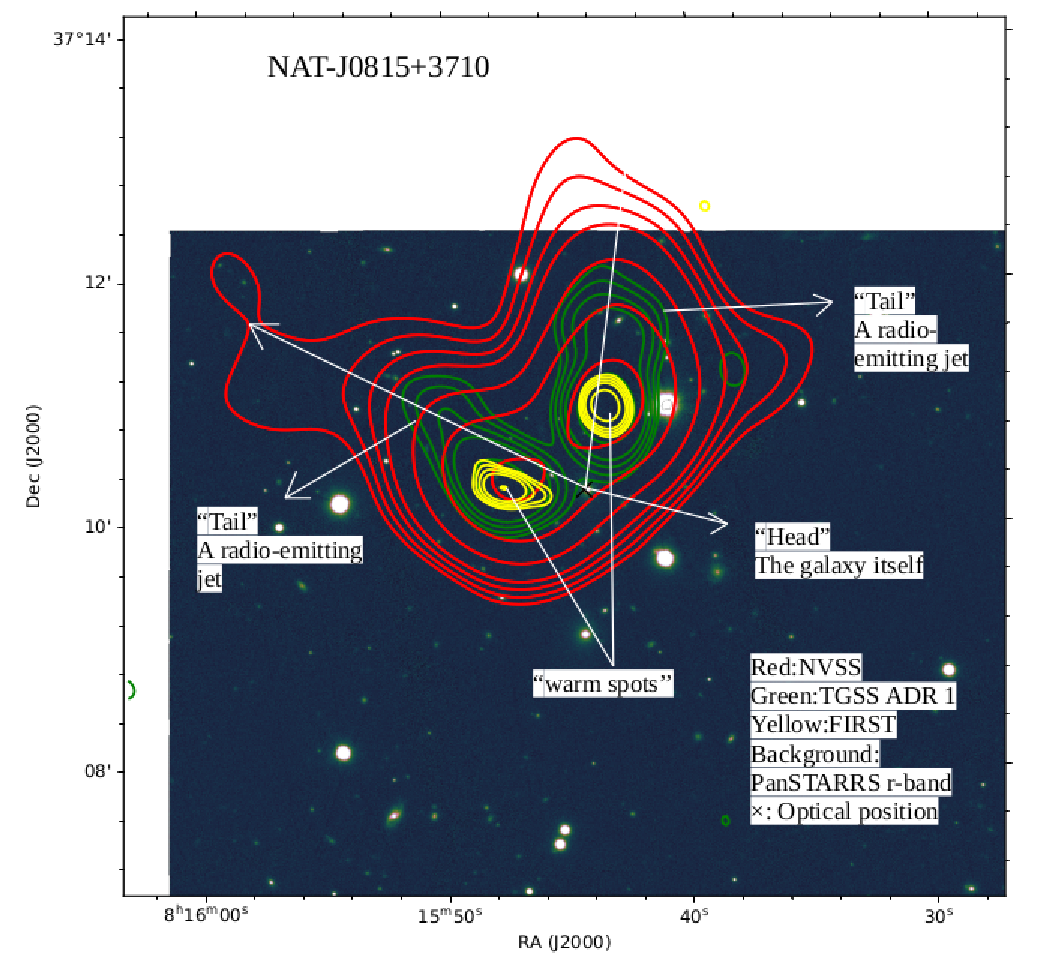}
\caption{A figure shows an example of ideal WAT (left) and NAT (right) sources. PanSTARRS r-band optical images are overlayed with TGSS, NVSS, and FIRST images. In both of the images, contour levels are at 3$\sigma \times$[-1, 1, 1.4, 2.0, 2.8, 5.6, 11.2, 22.4], where $\sigma$ for the TGSS, FIRST, and NVSS are 3.5, 0.15, and 0.45 mJy beam$^{-1}$, respectively. Both tailed radio sources are associated with host galaxies or radio cores (known as ``Head") that are indicated as cross marks. A pair of bright hotspots (called ``warmspots") are located closer to their radio core with respect to FR-II. The surface brightness is relatively higher towards the end of the jets than in the regions close to the core. Beyond these warmspots, well-collimated jets suddenly flare into diffuse plumes that may be significantly bent (known as ``Tails"). For the WAT, the angle ($\theta$) between two tails of the radio galaxy is $90\le \theta \le 180$ degrees, and for NAT, the angle is $\le 90$ degrees.}
\label{fig:ideal-WAT-NAT}
\end{figure*}

\subsection{Definition of WAT and NAT}
WATs are usually irregular radio galaxies with bright hotspots (known also as warm-spots with respect to FR-II) settled with host clusters, and their relative rest frame velocities are low ($<$ 1000~km~s$^{-1}$). Twins of well-collimated radio jets make obtuse bending angles ($\ge$ 90 degrees) under the influence of ICM weather/wind or induced merger shocks. In the 3C catalogue, WAT-3C 465 is known as its prototype \citep{Bur81, Elk84, Ei02, Ha05}. Because two tails are bent in the same direction, it is probably implied that the intra-cluster environment should have some regular ICM motion relative to the jets, rather than turbulence motion. We have included a source in our WAT sample if the bending angle is less than 160 degrees and if the source follows the selection criterion mentioned in subsection \ref{subsection:search}.

NATs are common in clusters, usually moving at high speeds through their central clusters ($\ge$ 2000 km~s$^{-1}$; \citet{Hu99, Su05}), and their distinctive structures reveal strong interactions with the inter-cluster medium, with radio tails bent in the same direction as core galaxies.

\subsection{Search from TGSS ADR 1}
\label{subsection:search}
We have made use of the TGSS ADR 1 consisting of images of a total of 5536 fields with a typical noise of $\sim$5 mJy at 150 MHz. The observations corresponding to this data release were executed between 2010 and early 2012 and cover about 90 percent of the whole sky. The survey covers 1 steradian of the southern sky, the maximum accessible southern sky from the observatory location. In search of tailed radio galaxies, we closely examine each of these 5536 image fields using Astronomical Imaging Processing Software ({\tt AIPS}). The good sensitivity and high resolution of TGSS (average rms of 3.5 mJy beam$^{-1}$) helps to study new fainter samples of different types of radio galaxies, such as X-Z shaped radio galaxies \citep{Bh20}, and giant radio galaxies \citep{Bh21a}. We visually examined all 5536 fields of TGSS images. For the selection of tailed radio source samples from TGSS, we tried the rigorous and tedious way and checked each of 5536 fields manually.
Our source selection criteria are as follows:

1. To make an initial list of probable WATs and NATs we make a list of radio sources showing two-sided jets with a clear bending angle $\le 160$ degrees.

2. The majority of tailed radio sources show a bright radio centre near host galaxies, which is known as ``head''. So, we go through all probable tailed radio sources to check if there is any bright source near the radio core location. To ensure radio core positions, we also compute the spectral index near the host galaxy and check whether the corresponding spectral index is flat. This radio core position is indicated as the cross mark.

3. For WAT/NAT sources, radio jets extend into plumes beyond the radio head. We identify sources with warmspots near both sides of the radio head. We measure the peak flux of warmspots of probable WAT/NAT sources, which should be higher than the tail radio flux.

4. We measure the angle between the two tails of all radio sources with the radio core. If the angle of the sources is greater than 90 degrees, then the source is catalogued as WAT, and if the angle is $\le 90$ degrees, the source is named as NAT.

5. All selected sources have a greater angular size than the synthesised beam ($\sim$4).

In figure \ref{fig:ideal-WAT-NAT}, we present the above mentioned criteria of an ideal radio image of tailed radio sources with surface brightness contours from TGSS, FIRST, and NVSS radio maps.

We excluded 20 bent sources from our list as they may be radio relics of a cluster. These possible relics are chosen using the following criterion. 

1. These diffuse and bent sources are located within 20 kpc of the position of the centre of a known galaxy cluster.

2. There are no optical/IR counterparts or radio cores in these sources. They also lack any warm-spots.

3. We study multi-wavelength radio images of these diffuse radio sources.

4. The power law radio spectrum is steeper ($\alpha \sim 1.2-1.3$).

Similar studies of tailed radio sources were recently conducted using a variety of methodologies. Double radio sources were found with different morphological radio sources (15 types) via automated pattern recognition using NVSS \citep{Pr11}. In this catalogue, 199 C-shaped radio galaxies (either WATs or NATs) were present. From the Proctor sample, \citet{Yu19} confirmed 412 C-shaped sources.

{\color{blue}The sample of detected WATS/NATS is flux limited and not complete as the detection depends on the lower contour of each image. It is complete upto 10.2 mJy beam$^{-1}$.}

\section{Counterpart identification}
\label{subsection:counterpart}
This section focuses on the identification of the optical and IR counterparts of the tailed radio sources. We use a likelihood ratio (LR) technique that is particularly useful when dealing with deep optical images to minimise the number of spurious associations. For 261 of our tailed radio sources, we find a reliable counterpart.
\subsection{Likelihood ratio technique}
The Likelihood ratio technique (LR; \citet{Ri75, De77, Su92, Ci03}) method allows us to take into account not only the position of the counterpart but also the background source magnitude distribution and the presence of multiple possible counterparts for the same radio source. It is given by the relationship \citep{Su92}
\begin{equation}
LR=\frac{q(m)f(r)}{n(m)}
\end{equation}
n(m) represents the surface density of background sources as a function of band magnitude $m$.This surface density is defined as the ratio of the magnitude distribution of background sources to the total searching area. The parameter $q(m)$ represents the {\it a priori} probability that the radio source has a counterpart of magnitude $m$. According to \citet{Ci03}, the parameter $q(m)$ is derived from
\begin{equation}
q(m)=\frac{real(m)\times Q}{\sum_{m_{i}}real(m)_{i}}
\end{equation}
The $real(m)$ is calculated from the following equation.
\begin{equation}
real(m) = total(m)-n(m)\times N_{radio}\times \pi r_{max}^{2}
\end{equation}
where $N_{\textrm{radio}}$ is the number of radio sources in the catalogue, $total(m)$ is the objects in the subsidiary catalogue within a radius of $r_{max} (\sim 2''$) around each radio source and the $Q$ is usually estimated by determining the fraction of sources with radio counterparts above the background as follows:
\begin{equation}
Q=\frac{N_{\textrm{counterparts}}-(\sum_{m_{i}} n(m)\times r_{max}^{2}\times N_{\textrm{radio}})}{N_{\textrm{radio}}}
\end{equation}
The parameter $f(r)$ represents the probability distribution of the off-set $r$ between the catalogued positions of the radio source and its potential counterpart. The uncertainty in this offset is calculated by combining the uncertainty in the radio position, the uncertainty in the optical/IR position, and the uncertainty in the relative astrometry of the two surveys. As for $f(r)$, we adopt a two-dimensional Gaussian distribution of the form:
\begin{equation}
f(r)=\frac{1}{2\pi \sigma^{2}}\exp(\frac{-1}{2{\sigma}^{2}})
\end{equation}
For each source, $\sigma$ is the average value between $\sigma_{x}=\sqrt{er_{op}^{2}+\sigma_{\alpha}^{2}}$ and $\sigma_{y}=\sqrt{er_{op}^{2}+\sigma_{\delta}^{2}}$, where $er_{op}$ is the position error of the counterpart and $\sigma_{\alpha}$ and $\sigma_{\delta}$ are the radio positional errors in RA and DEC.
The LR does not contain information about the possible presence of many counterpart candidates in the surrounding of a specific radio source. It is therefore useful to define the reliability of each association as:
\begin{equation}
\rho_{j}=\frac{(LR)_{j}}{\sum_{i}(LR)_{i}+(1-Q)}
\end{equation}
where the sum is over all the candidate counterparts for the same radio source \citep{Su92}.
\subsection{Subsidiary catalogue of WISE and Pan-STARRS data}
The selection of multi-wavelength data is crucial to identify the correct counterpart of a radio source. We use two deep optical surveys from the Panoramic Survey Telescope and Rapid Response System (Pan-STARRS; \citet{Ch16}) and infrared survey from the Wide-field Infrared Survey Explorer (WISE; \citet{Wr10}). Deep and wide optical and IR data are available over the TGSS-ADR 1 covered sky. Pan-STARRS1 has performed a set of independent synoptic imaging sky surveys, including the $3\pi$ steradian survey. This survey covers the entire northern sky and southern sky upto $\delta >-30$ degrees. There are five bands present $grizy$ (23.3, 23.2, 23.1, 22.3, 21.4 mag). The typical point spread function (PSF) of the Pan-STARRS images is $\sim 1-1.3''$. In the astrometric calibration, the uncertainty of the standard deviation of the mean and median residuals ($\Delta$ra, $\Delta$dec ) are (2.3, 1.7) milliarcsec and (3.1, 4.8) milliarcsec, respectively.
The WISE covers a mid-infrared survey of the entire sky. The sensitivity of this survey is much higher than previous infrared survey missions. WISE achieved a sensitivity more than 100 times better than IRAS in the 12 $\mu$m band. WISE is covering the whole sky in four infrared bands W1, W2, W3, and W4 in the 3.4, 4.6, 12, and 22 $\mu$m respectively. The All WISE catalogue includes more than 747 million sources across the full sky. The W1 and W2 bands have significantly better sensitivity than the other two WISE bands \citep{Cu13}. The completeness of the WISE catalogue varies across the sky.With W1 $<$ 19.8, W2 $<$ 19.0, W3 $<$ 16.67, and W4 $<$ 14.32 mag, the completeness is 95 per cent for sources.
Initially, we included all WISE sources smaller than 15$''$. The WISE W1-band and Pan-STARRS $i$-band data sets are combined using LR band magnitude analysis. For each WISE source, we searched for the best Pan-STARRS match in $i$ band. LR ratios are then derived for all PanSTARRS sources within 15$''$ of all WISE position. And for each WISE source, the highest LR above the threshold limit is taken as the PanSTARRS counterpart. Finally, we make a WISE-Pan-STARRS combined catalogue that should be used for counterpart identification.
\subsection{Cross-match of tailed radio sources with subsidiary catalogues}
We use the magnitude and colour in the LR method to cross-match the TGSS ADR1 tailed radio sources with the subsidiary WISE-Pan-STARRS combined catalogue. Following the LR method, we built a list of possible counterparts for each of the tailed radio sources. Initially, we set a very low likelihood threshold to be sure not to lose any counterparts. After a careful analysis, we select only those sources with a reliability greater than 0.6, as the threshold limit to ensure the expected number of spurious associations is 4 per cent of the subsidiary catalogue. As a result, maximising the number of identified tailed radio sources was maximised. Another approach to measuring a value for the reliability limit, where those sources with a reliability greater than $\rho_{c}$ can be accepted as true counterparts, was done by Smith et al. 2011. They estimated the number of false cross-matches using $N_{false}=\sum_{\rho \ge0.8} (1-\rho)$ and determined the contamination rate $\sim$4.2 (ratio of $N_{false}$ to matched sources with reliability limit). In our catalogue, the number of $N_{\textrm{false}}$ is 12 and matched sources with reliability limit is 261. The contamination rate is 4.5 per cent with the reliability limit of 0.6 using the above relation.

\begin{figure}
    \centering{
\includegraphics[width=6.0cm,angle=270,origin=c]{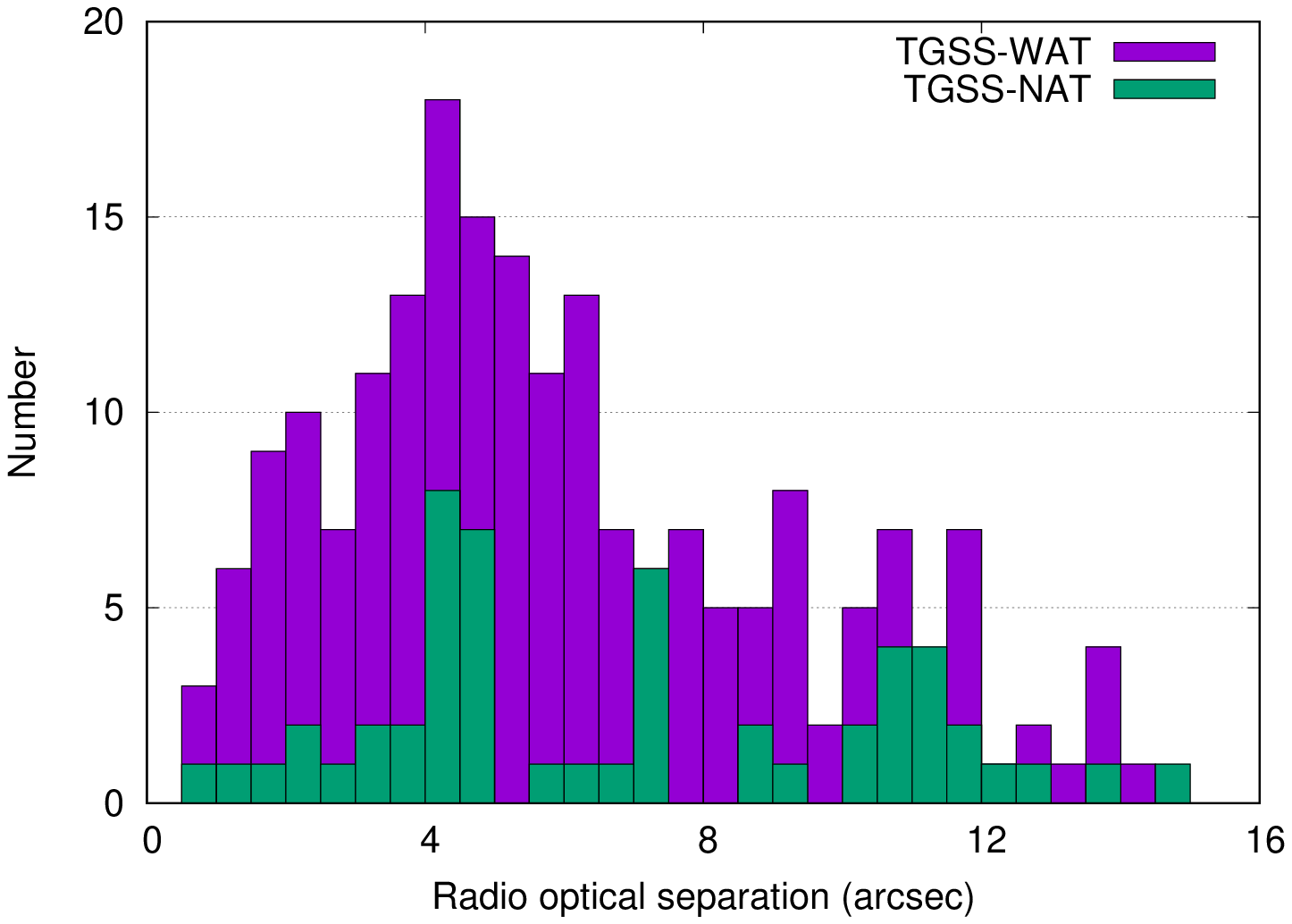}
}
\caption{Distribution represent the separation between radio and optical positions of tailed radio sources.}
\label{fig:radio-optical}
\end{figure}

The choice of the best threshold value $LR_{thr}$ is necessary to discriminate between spurious and real identifications. Here, $LR_{thr}$ should be large enough to keep the number of spurious identifications as low as possible and to increase the reliability. We defined a source as counterparts tailed radio sources if their LRs are above the thresholds of $LR_{thr}$ = 7.57 in the i-band or $LR_{thr}$ = 0.77 in the W1-band. If more than one spurious counterpart is above those thresholds, then the counterpart with the highest LR in either of the two bands is accepted and the other is discarded. The counterpart identification rates in our catalogue are 61 per cent and 98 per cent in the $i$ and W1 bands, respectively. We also calculate the radio optical separation of each radio source with the highest reliability. The distribution of radio optical separation is shown in Table \ref{tab:WAT} and Table \ref{tab:NAT}. With this criteria, a 3 per cent counterpart (8/261) is identified with spurious association.
Figure \ref{fig:radio-optical} presents the distribution of separation between optical and radio counterparts. A peak is seen near 4 arcsec.

We make use of other previous all-sky surveys. In particular, we use the SDSS DR-12 catalogue \citep{Al15} and the Two Micron All Sky Survey (2MASS; \citet{Sk06}) extended source catalogue (2MASX; \citet{Ja00}). We found the optical counterparts of 57 WATs in SDSS, 26 in the 2MASS catalogue, and 13 in the 2MASX catalogue, respectively. We also identify optical counterparts for 10 NATs in SDSS, 11 in the 2MASS catalogue, and 17 in the 2MASX catalogue, respectively. In our catalogue, about 117 tailed radio galaxies (92 WATs and 25 NATs) have redshift information. For 92 WATs, the spectroscopic redshift data that can be collected from the SDSS is 34, for NATs it is three.
Since optical/IR counterparts are more compact than the corresponding radio galaxies, we used the position of optical/IR counterparts as the position of these sources. For the rest of the 4 (1 per cent of total detected sources) tailed radio sources, where optical or IR counterparts are not available, a radio-morphology based position is used.

By visual search, \citet{De14} detected 56 bent tailed radio sources over $\sim$4 square degrees area of the southern sky in the ATLAS field. From combined mosaic radio sources (FR-I and FR-II) of the southern sky, new 24 bent tailed samples were identified by visual inspection \citep{Ob18}. \citet{Mi19} presented a catalogue of 47 wide-angle tailed radio galaxies (WATs). All candidates were selected by combining observations from the NVSS, FIRST, and SDSS surveys. With the help of the VLA FIRST survey at 1.4 GHz, \citet{Pa21a} found 614 new head-tail radio galaxies, among them, 398 were WATs and 216 were NAT sources. \citet{Pa21} presented a catalogue of fifty new head-tail radio sources (five NATs and forty-five WATs) using  LOFAR Two-metre Sky Survey first data release (LoTSS DR1) at 144 MHz frequency \citep{Sh19}. The survey coverage area of the present paper is larger than all other previous works.

\section{Results}
\label{sec:result}
Different information about the objects reported in this paper is given in Table \ref{tab:WAT} and Table \ref{tab:NAT}. In the first two columns, the catalogue number and identification names are given. Columns (3) and (4) contain the J2000 coordinates of the optical/IR counterpart. The radio counterpart separation is presented in column (5). When the optical counterpart is not found, the approximate position using the morphology of the radio source is provided. The reliability of counterpart ($\rho$) is presented in columns (6). In columns (7) and (8), the total flux density in Jy at $150$ MHz ($F_{150}$) and $1400$ MHz ($F_{1400}$) obtained from the TGSS and NVSS surveys is provided. Columns (9) and (10) contain the spectral index and redshift of the sources respectively. In column (11), we provide the luminosity in 150 MHz. Column (12) contains the names of earlier radio surveys where the source is presented without identification of them as tailed radio galaxies.

\begin{figure}
    \centering{
\includegraphics[width=8cm,angle=0,origin=c]{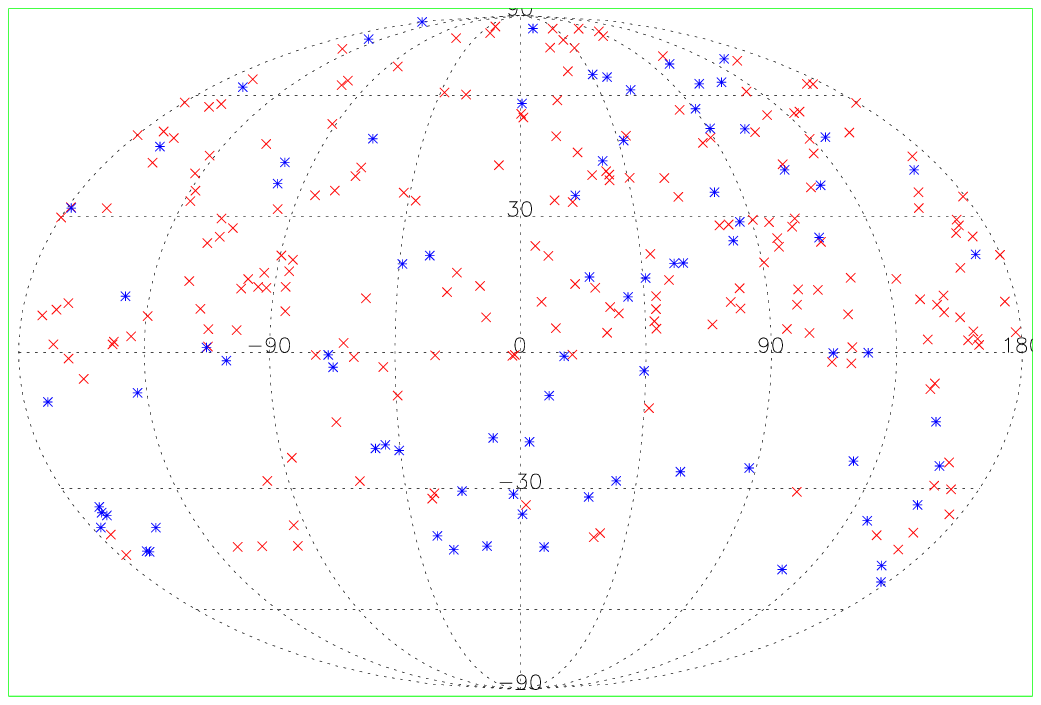}
}
\caption{The spatial distribution of tailed radio sources. The `$\times$' symbols indicate WAT sources and the `$\ast$' symbols represent NAT sources.}
\label{fig:source-distribution}
\end{figure}
We report the discovery of 203 WAT and 61 NAT sources from the TGSS ADR 1 at 150 MHz. We remove sources reported in \citet{Pr11} and other catalogues from our list \citep{De14, Mi19, Yu19}. \citet{De14} have identified 45 bent-tailed (BT) radio galaxies in Australia Telescope Large Area Survey (ATLAS) at 1.4 GHz, which is supplemented with the 1.4 GHz Very Large Array images. \citet{Pr11} identified 199 WATs or NATs radio galaxies by using an automated morphological classification scheme. \citet{Mi19} presented a catalogue of 47 wide-angle tailed radio galaxies (WATs). All candidates were selected by combining observations of the NVSS survey, FIRST survey, and SDSS survey. We present the spatial distribution of the newly identified WAT and NAT sources in Figure \ref{fig:source-distribution}. As expected, the sources show a random distribution. Many of these newly identified sources are catalogued as two separate sources in the NVSS catalogue \citep{Con98}. 
The number of sources is higher in the northern region compared to the southern region because the survey is limited up to a declination of --55 degrees and the local RMS in the southern hemisphere is higher compared to that of the northern hemisphere due to high declination effects.

We present high-sensitivity 150 MHz GMRT images of ideal selected WAT and NAT radio galaxies Figure \ref{fig:ideal-WAT-NAT} identified from the TGSS as examples of WAT and NAT sources discovered under the present project. We present the radio image of tailed radio sources with different contours from TGSS ADR 1, NVSS, and FIRST. PanSTARRS r-band optical images are overlayed with TGSS, NVSS, and FIRST images. In all figures, contour levels are at 3$\sigma \times$[-1, 1, 1.4, 2.0, 2.8, 5.6, 11.2, 22.4], $\sigma$ for the TGSS, FIRST, and NVSS are 3.5, 0.15, and 0.45 mJy beam$^{-1}$, respectively. Overlayed images of all the tailed radio sources are available as supplementary material in the online version of the article.
\subsection{Spectral index}
\label{subsec:spectralindex}
The two-point spectral index of newly discovered radio galaxies between 150 and 1400 MHz is calculated assuming $F \propto \nu^{-\alpha}$, where $\alpha$ is the spectral index and $F_{\nu}$ is the radiative flux density at a given frequency $\nu$. These spectral indices have been determined by integrating fluxes over the same aperture at both frequencies,  using formula
\begin{equation}
 \alpha= \frac{\log F_{{\nu}_{1}}- \log F_{{\nu}_{2}}}{\log \nu_{2}-\log \nu_{1}}
\end{equation}

The integrated radio flux density of tailed-radio sources can be estimated using Astronomical Imaging Processing Software (AIPS) with {\tt TVSTAT}. To measure the flux density in TGSS and NVSS field, for uniformity, we use a region corresponding to the 10.5 mJy beam$^{-1}$ ($3\sigma)$ contour level from TGSS. 

\begin{figure*}
\includegraphics[width=6cm,angle=270,origin=c]{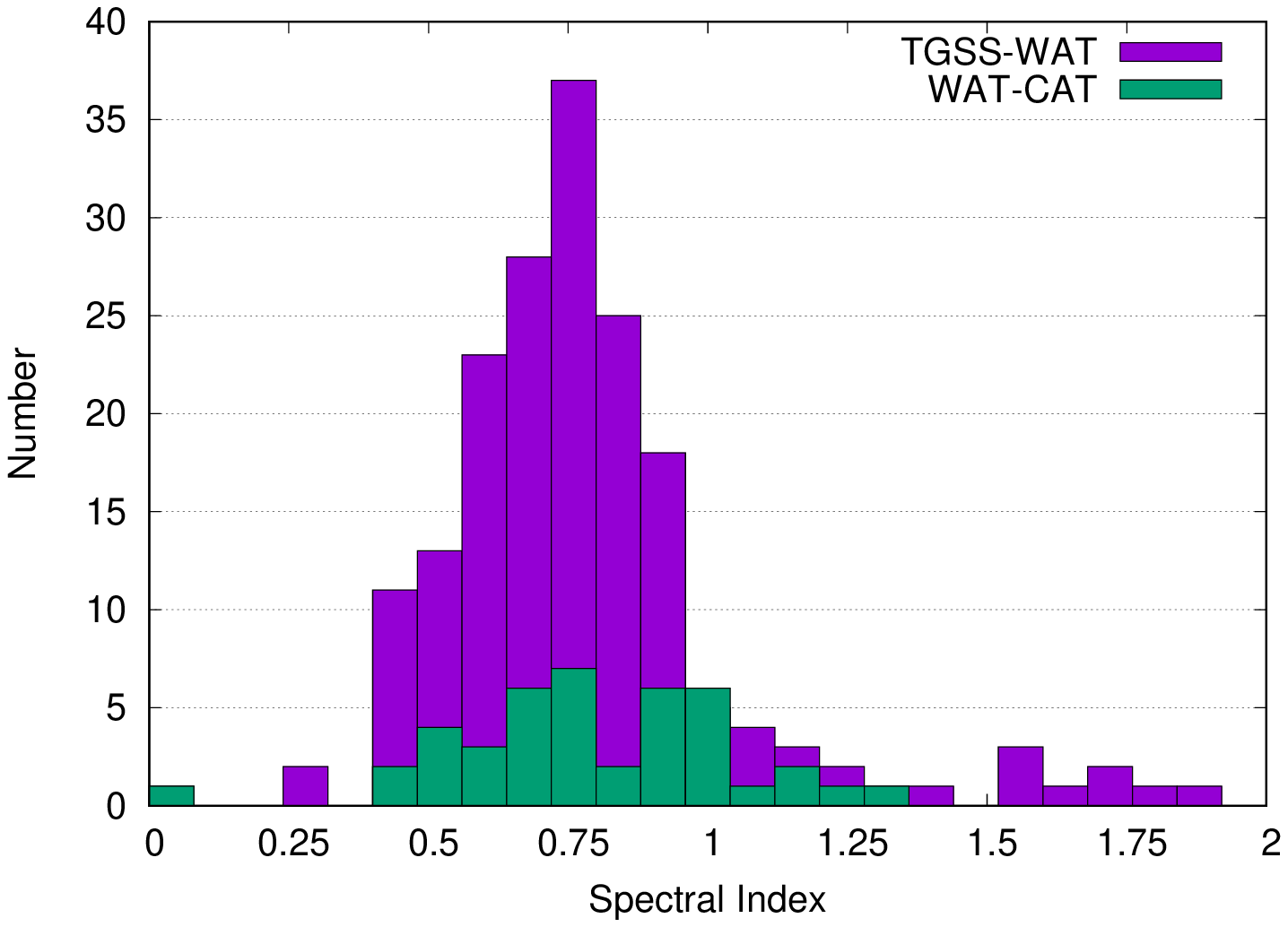}
\includegraphics[width=6cm,angle=270,origin=c]{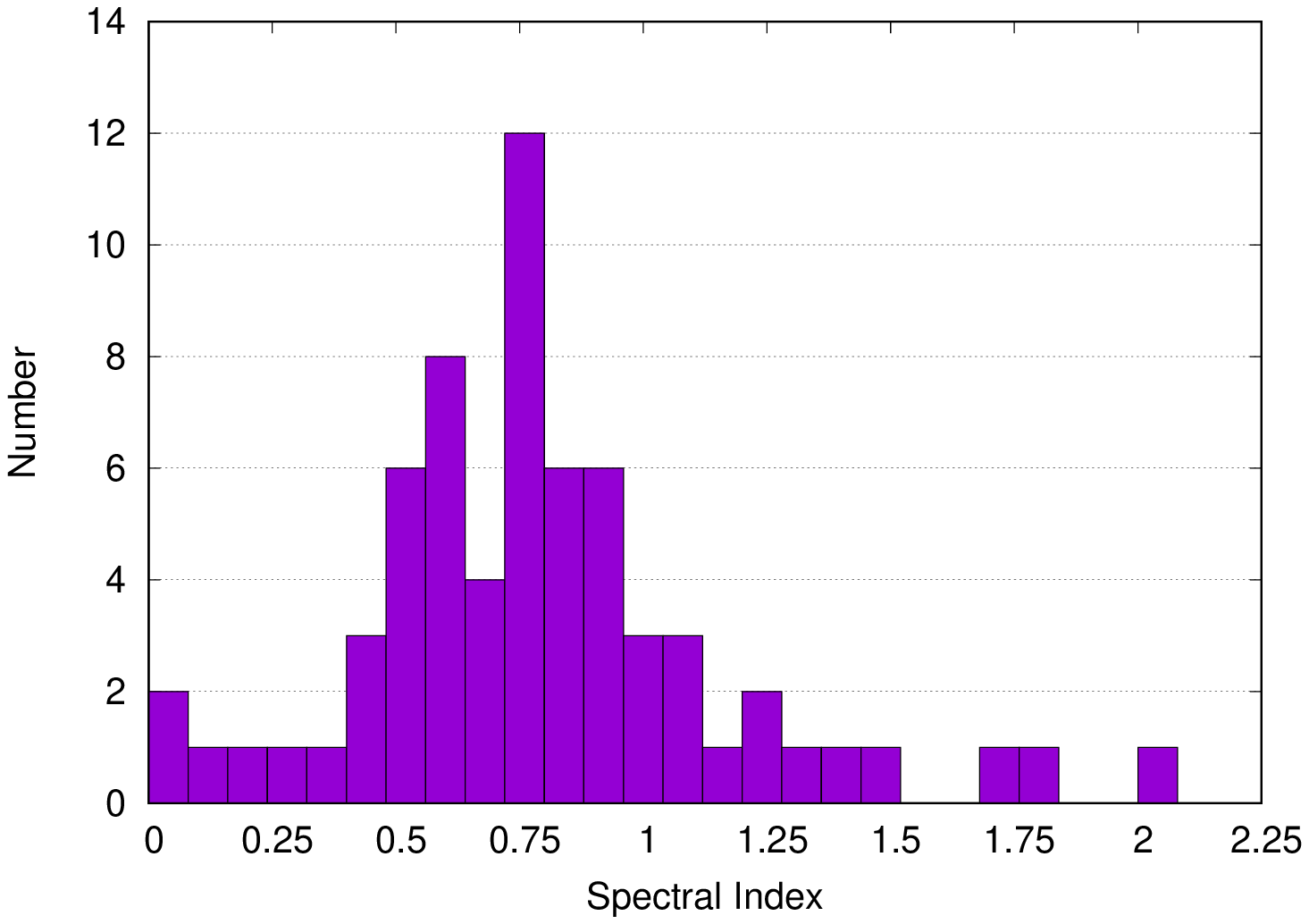}
 \caption{Distribution showing spectral index distribution for WAT (left) and NAT (right) sources. We also included sources presented in WAT-CAT catalogue \citet{Mi19}.}
\label{fig:alpha-distr}
\end{figure*}

The spectral index ($\alpha_{150}^{1400}$) is mentioned in Table \ref{tab:WAT} and Table \ref{tab:NAT}. For 188 WATs and 51 NATs, the spectral index measurements are available. The remaining 15 WATs and 10 NATs were not detectable in NVSS maps since their declination is less than the NVSS coverage. Out of 188 WATs with spectral index information, 13 (7 per cent) show flat spectrum ($\alpha_{150}^{1400}<0.5$). Out of 61 NATs with spectral index information, 5 (7 per cent) are showing flat spectrum ($\alpha_{150}^{1400}<0.5$). Most of the WATs and NATs show a steep radio spectrum ($\alpha_{150}^{1400}>0.5$) which is a common property of lobe dominated radio galaxies.
The uncertainty of spectral index measurements due to flux density uncertainty \citep{Mah16} is
\begin{equation}
\Delta\alpha=\frac{1}{\ln\frac{\nu_{1}}{\nu_{2}}}\sqrt{\left(\frac{\Delta S_{1}}{S_{1}}\right)^{2}+\left(\frac{\Delta S_{2}}{S_{2}}\right)^{2}}
\end{equation}

where $F_{\nu_{1, 2}}$ and $S_{1, 2}$ refers to NVSS and TGSS frequencies and flux densities respectively. The flux density accuracy in TGSS ADR 1 and NVSS is $\sim10$ per cent \citep{Int17} and $\sim5$ per cent \citep{Con98}. Using equation 2, the spectral index uncertainty is $\Delta\alpha$=0.05.
The spectral index distribution for WATs (left) and NATs (right) with sources presented in the current article is shown in Figure \ref{fig:alpha-distr}. The distribution shows slightly different peaks for WATs and NATs. The distribution peaks near $0.6-0.7$ for WATs and near $0.65-0.75$ for NATs. For WAT, the total span of $\alpha_{150}^{1400}$ is from $0.02$ to $1.86$ and for NATs, the total span of $\alpha^{1400}_{150}$ is from $0.38$ to $2.04$. Among WATs, J0752+0814 has the highest spectral index (with $\alpha^{1400}_{150}$=1.70) and J0041+2104 has the lowest spectral index (with $\alpha^{1400}_{150}$=0.02). For NATs, J0704+6318 has the highest spectral index (with $\alpha^{1400}_{150}$=2.04) and J2144+2107 has the lowest spectral index (with $\alpha^{1400}_{150}$=0.38). 
The radio spectral index is commonly utilised to differentiate between source components like core and lobe. To ensure the location of the core, we compute spectral index of the probable core region (found from optical/IR counterpart and radio morphology) using TGSS and FIRST survey (when available) and found that all of them are showing flat spectral index ($\alpha_{150}^{1400}<0.5$). The spectral index of the core is shown in images of all sources, when corresponding FIRST images are available.

\subsection{Luminosity feature}
\label{subsec:lum}
\begin{figure}
\centering{
\includegraphics[width=6cm,angle=270,origin=c]{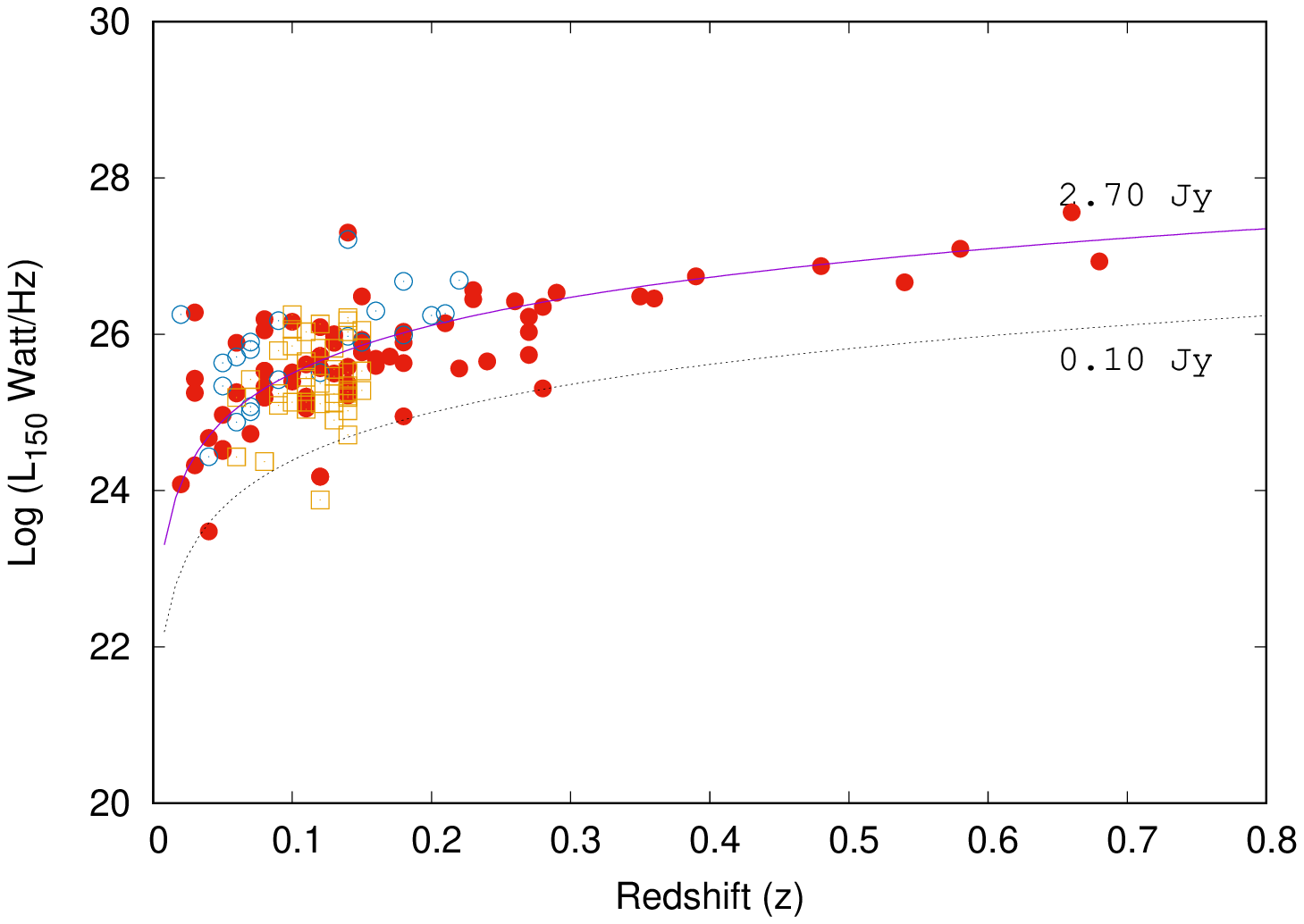}}
\caption{Plot of radio luminosity against redshift for the 103 tailed sources presented in the current paper. The WAT and NAT sources are indicated by red circles and open blue circles, respectively. The yellow squares present wide-angle tailed radio galaxies in the WATCAT catalogue \citep{Mi19}. The pink solid line indicates the best-fitted luminosity using points from all surveys in the figure, which corresponds to 2.70 Jy flux density for tailed radio sources.}
\label{fig:lum}
\end{figure}

We have calculated the radio luminosities ($L_{150}$) of sources using the redshift information of either the optical counterpart or the associated galaxy cluster using the standard formula \citep{Do09}
\begin{equation}
L_{150} = 4\pi D_{L}^{2}S_{0}(1+z)^{\alpha-1}
\end{equation}
where $z$ is the redshift parameter, $\alpha$ is the spectral index ($S \propto \nu^{-\alpha}$), $D_{L}$ is luminosity distance of the source (Mpc), and $S_0$ is the flux density (Jy) at $150$ MHz frequency.
Figure \ref{fig:lum} shows the redshift vs luminosity plot of the $103$ tailed sources for which either spectroscopic or photometric redshift is available. In our sample, the highest redshifts for the NAT and WAT sources are $0.51$ and $0.68$ respectively.
The source radio luminosities at 150 MHz are in the order of $10^{25}$ W Hz$^{-1}$, which is similar to a typical radio galaxy.
The average value of $Log ~L$ [W Hz$^{-1}$] for WATs is 25.62 (1$\sigma$ standard deviation$=0.72$, median$=25.63$) and that of NATs is 25.82 (1$\sigma$ standard deviation$=0.72$, median$=25.83$). 
J0856+4829 is the least luminous WAT in our sample with $L_{150}=0.15\times10^{25}$ W Hz$^{-1}$ ($z=0.12$) and J0549--2520 is the least luminous NAT in our sample with $L_{150}=0.27\times10^{25}$ W Hz$^{-1}$ ($z=0.04$). J0225+4031 is the most luminous WAT in our sample with $L_{150}=1.9 \times10^{27}$ W Hz$^{-1}$ ($z\sim0.65$) and J1314+6220 is the most luminous NAT in our sample with $L_{150}=162.8\times10^{25}$ W Hz$^{-1}$ ($z=0.14$). 
\subsection{Bending angle}
\label{subsec:bending}
\begin{figure}
\centering{
\includegraphics[width=7.5cm,origin=c]{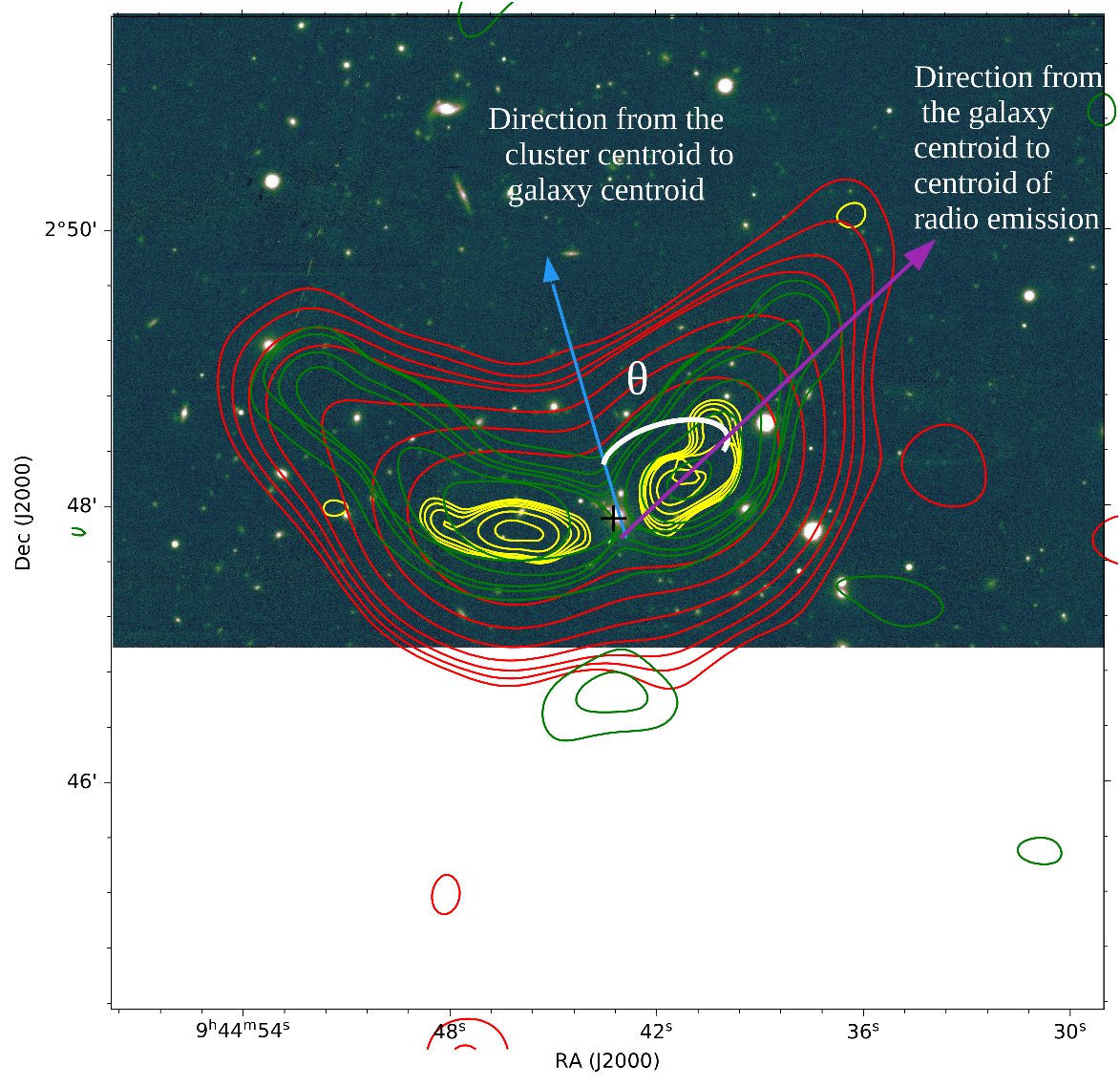}}
\caption{A figure showing an example of a measurement of the bending angle of a tailed radio source. PanSTARRS r-band optical images are overlayed with TGSS (green contour), NVSS (red contour), and FIRST (yellow contour) images. The '+' symbol is the position of the host galaxy from PanSTARRS r-band where two radio tails are terminated. The blue arrow indicates the direction from the cluster centroid to the galaxy centroid, and the magenta arrow shows the direction from the galaxy centroid to the centroid of the radio emission peak in the intensity map. We calculate the angle $\theta$ between two arrows. Then 2$\theta$ is the bending angle between two radio tails.}
\label{fig:bend-angle}
\end{figure}

\begin{figure}
\centering{
\includegraphics[width=6cm,angle=270,origin=c]{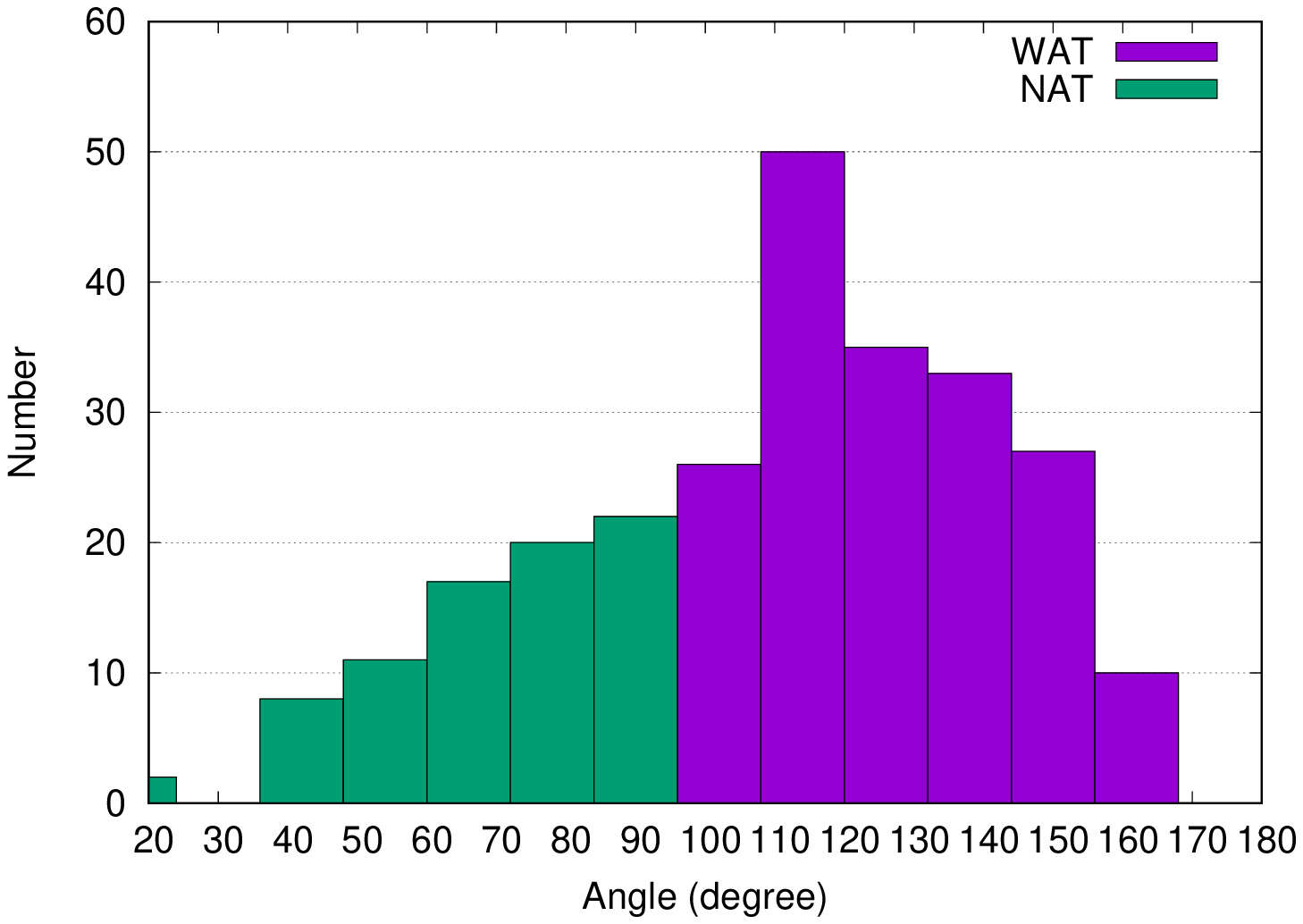}}
\caption{Distribution showing source count vs bending angle of the tailed galaxies. The green and blue bars indicate the number of NAT and WAT sources respectively in the range of a particular bending angle.}
\label{fig:angle}
\end{figure}

We measure the bending angle of all sources using the angle (2$\theta$) between two individual jet axis connecting the core that is shown in Figure \ref{fig:bend-angle}. We classified all tailed sources in our sample into two groups -- `wide-angle tailed' galaxy or `narrow-angle tailed' depending on the bending angle of these sources. For some sources, the bending angle could not be measured because of the complex structure of these sources.
Figure \ref{fig:angle} presents the distribution showing the distribution of bending angles of sources presented in the current paper for both NAT and WAT type sources. Most of the NAT sources made an angle greater than $80\degr$. WAT sources produced a wide variety of angles between the two components, with a peak near 110--120$\degr$.

\subsection{Cluster Association}
\label{subsec:cluster}
In the Universe, clusters are the greatest gravitationally bound structures \citep{We12}. Clusters give an ideal tool for cosmologists to understand the composition and evolution of the structures of the Universe \citep{Al11, Wet12}. They provide strong proof for the appearance of large amounts (80 per cent) of dark matter \citep{We12}. They are created around dark matter concentrations when two cosmic objects (sheets and filaments) are intersected. Understanding clusters is very helpful for tracing the large-scale structure of the Universe \citep{Pr88, Hi91, Wo00, Be07, Ta08}.
We associate our tailed radio galaxy sample with cluster catalogues from the literature that covers the TGSS field. These clusters have been detected using various methods, including optical, IR, X-ray, and SZ observations. The cluster catalogues used are listed in Table \ref{tab:wat-cluster} and Table \ref{tab:nat-cluster}. Using only the 107 WATs and NATs candidate sources with redshifts, we performed a three-dimensional cross-match with the known clusters across the field using a search radius of 2 Mpc. The distance is computed using the source redshift. We assume that tailed-radio galaxies and clusters of galaxies are associated if $\Delta z= \mid z-z_{spec} \mid \le 0.005$, where $z_{spec}$ is the spectroscopic redshift of galaxy cluster \citep{Mo93, Ek04, Be06}.
We found associated known clusters or groups for 120 tailed radio galaxies from our WATs and NATs sample (out of 264 total). The details of associated clusters for WATs and NATs presented in the current paper are listed in Table \ref{tab:wat-cluster} and Table \ref{tab:nat-cluster}. In columns (1) and (2), the catalogue number and galaxy cluster identification name are given. In columns (3) and (4) the name of the catalogue where the cluster is named and the redshift of the cluster is given. Column (5) and (6) contains the comoving distance ($D_{c}$) in Mpc at sources redshift and angular separation (in ars) between the centre of the associated cluster and galaxy centre. Column (7) indicates the linear distance of the host galaxy from the cluster centre. In column (8), the BCG r band magnitude $m_{r}$ is shown. In column (9) and (10), cluster radius ($r_{500}$) and cluster richness ($R_{L}$) is shown. In column (11), the number of member galaxy candidates ($N_{500}$) within $r_{500}$ is shown and in column (12), the mass of the cluster within $r_{500}$ ($M_{500}$) is presented. The cluster density ($\rho_{co}$) is presented in column (13). 
 For a homogeneous study, various parameters ($m_{r}$, $r_{200}$, $N_{500}$, and $R_{L}$) of clusters in Table \ref{tab:wat-cluster} and Table \ref{tab:nat-cluster} are collected from \cite{We12, We15}. We calculate the optical mass $M_{500}$ and cluster density $\rho_{co}$ using these parameters. We also found that for 65 tailed-radio sources in our sample, the distance between two sources is less than 500 kpc. Search for a new cluster of galaxies is encouraged near these WATs and NATs when no associated cluster has been identified.
 
\section{Discussion}
\label{sec:disc}
\subsection{Radio properties of WATs and NATs}

Based on our study, we further discuss our results of tailed radio sources in this section. In this section, we also include other samples (WAT-CAT catalogue, LoFAR catalogue) and compare with their results.

\begin{figure}
\includegraphics[width=6cm,angle=270,origin=c]{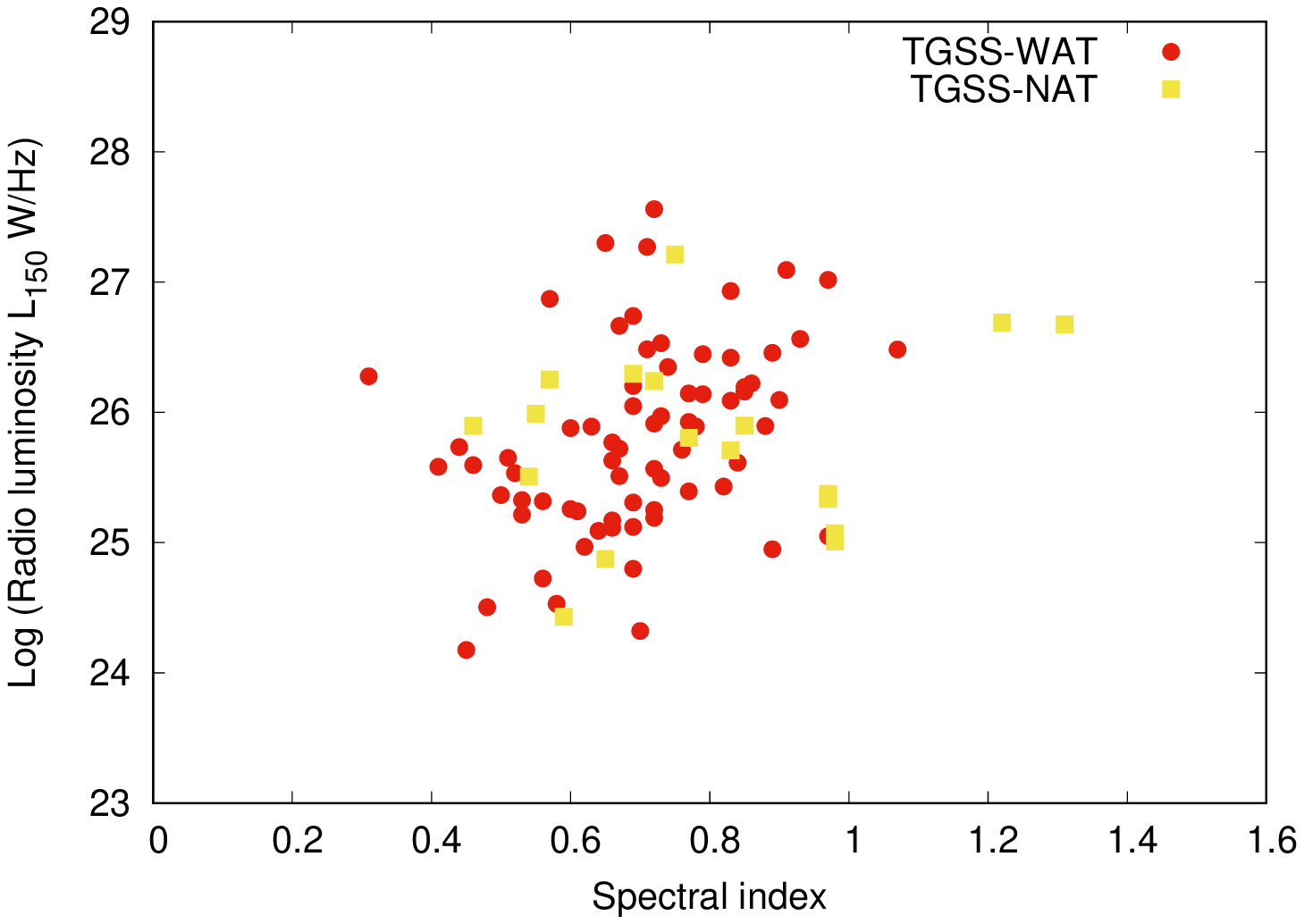}
\caption{Plot shows spectral index variation with radio luminosity of tailed radio galaxies.}
\label{fig:spectral-luminosity}
\end{figure}

In Figure \ref{fig:spectral-luminosity}, the variation of radio luminosity with the spectral index of tailed radio galaxies is shown. The figure indicates that the spectral index tends to increase with the radio power. The correlation is significant for our sample, which was selected at a low frequency of 150 MHz. This result is consistent with previous studies \citep{Bl99}. There is a tendency for the majority of powerful radio sources to have higher spectral index. At a low frequency, the spectral index gives us information on the energy index of the synchrotron particles that are injected into the lobes \citep{Bl99}. With the help of enhanced magnetic field, radio sources create high jet kinetic power \citep{Bl99}, which helps fast synchrotron cooling of relativistic electrons. The cooling time ($\tau$) varies with the magnetic field ($\tau \propto \frac{1}{B^{2}}$). At last, the magnetic field enhances the synchrotron loss and make the steeper energy distribution at high radio luminosity.

Figure \ref{fig:lum} represents the distribution of tailed radio sources in the $\log~L$--$z$ plane.  Most of the sources are within the redshift range of 0.1 to 0.4, and the number of radio sources decreases with increasing redshift beyond $z = 0.5$. There are no available sources in the lower right quadrant of Figure \ref{fig:lum} which is due to the non-identification of low luminosity radio galaxies at high redshift. This sensitivity limit of the survey is known as the Malmquist bias \citep{Ma22, Ma36}. The Malmquist bias means that there will be a tight correlation between luminosity and redshift in a single flux-limited sample. A tight correlation depends on (i) the steepness of the distribution in jet kinetic powers, (ii) the energy distribution of the particles injected into lobes \citep{Bl99}. This is expected because radio luminosity strongly correlates with the above two parameters. But, at high redshift, a negative correlation is established between projected linear size and redshift. 

Based on the availability of redshift data, we estimate $L_{150}$ for 79 WATs and 21 NATs from our catalogue. The mean and median $L_{150}$ of WATs and NATs confirm that WATs and NATs have the same distribution of the radio luminosity which implies that governing conditions in the core engines of WATs and NATs may be same. The mean and median of $Log ~L$ [W Hz$^{-1}$] of WATs in \citet{Mi19} are 25.40 and 25.35, and in \citet{Pa21} are 25.59 and 25.60 respectively, which indicate that our tailed radio galaxies are similar luminous compared to tailed radio galaxies in the FIRST and LOFAR catalogues.

The spectral index features of the tailed radio source population can be used to identify a variety of source characteristics. For example, the radio spectral index is commonly utilised to differentiate between the source components of AGN. The nature of core radio spectral indices is normally flat or inverted. Inverted radiation spectra probably originate in partially optically thick regimes where low frequency radiation is preferentially put down via free-free absorption or synchrotron self-absorption. On the other hand, flat spectra likely arise from the superposition of naturally distinct regions along the jet with dynamically lower frequency turnover. Steep energy spectrum usually comes from lobe components and ultra steep energy spectrum arise for the relic emission \citep{Mah16}. All previous studies \citep{Mah16, Is10, Ka98} found the mean spectral index of radio galaxies in the range of 0.7--0.8 and 0.75 are usually considered to be the typical spectral index for radio galaxies (RGs). For example, \citet{Mah16} found radio galaxies with spectral index in the range --0.5 to 1.5 with median=0.78 and \citet{Is10} found radio galaxies with spectral index in the range --0.5 to 2.5 with median=0.78. The mean and median of the spectral index of tailed radio sources presented in the current article is close to the spectral index of a typical radio galaxy which implies that tailed radio galaxies and normal radio galaxies are similar in terms of the properties of their spectral index.

\begin{figure}
\includegraphics[width=6cm,angle=270,origin=c]{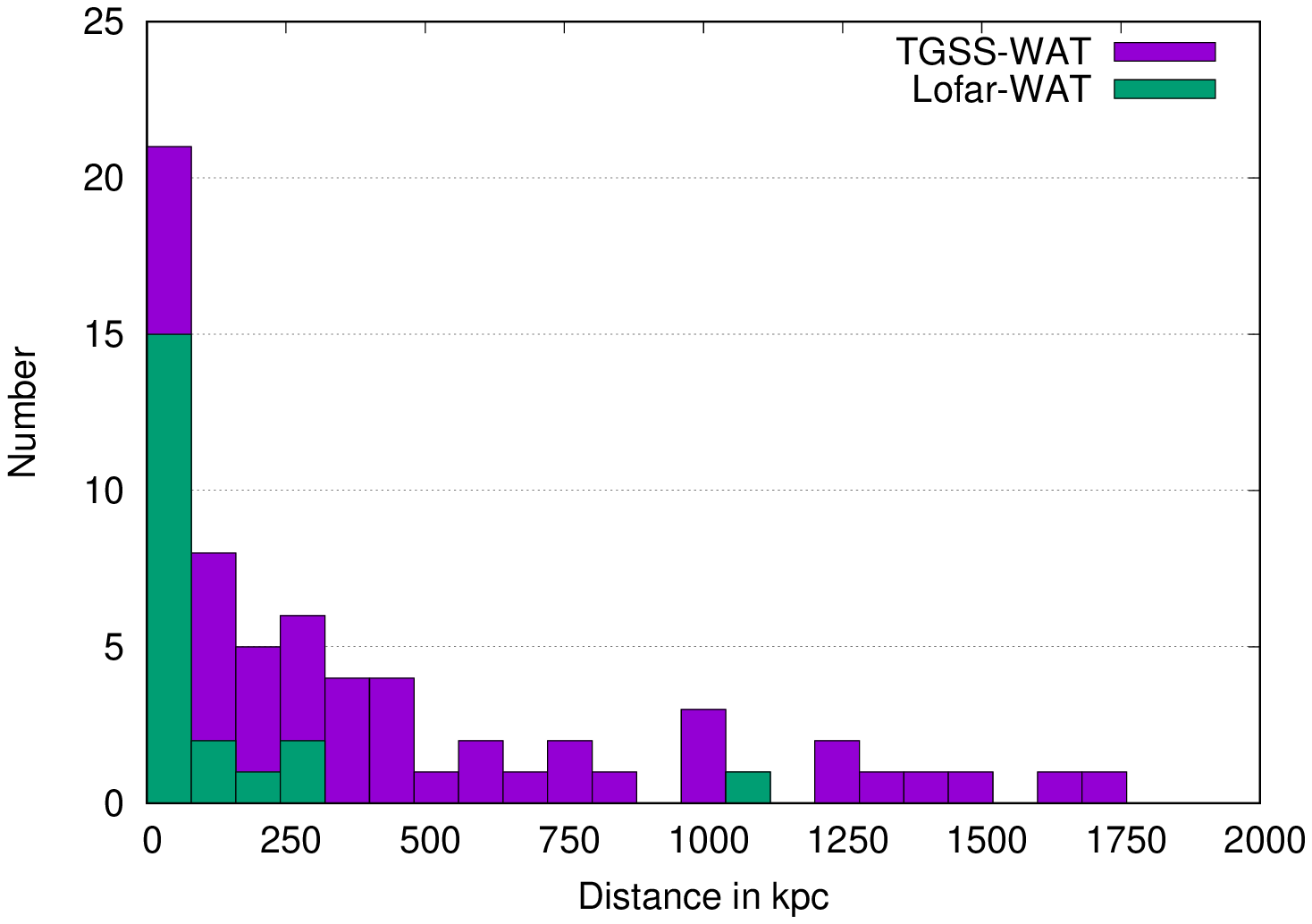}
\caption{Distribution of distance of tailed radio galaxies from the galaxy cluster centre. We also included sources presented in LoFAR WATs \citep{Pa21}.}
\label{fig:wat-spatial}
\end{figure}

Figure \ref{fig:wat-spatial} shows the distribution of distance of WATs with respect to the centre of host cluster. The cluster centric distance distribution peak is seen at distance $< 250$ kpc. For distances above $\sim$600 kpc the fraction of WATs drops quickly. This figure shows that the majority of the WATs are lying close to the cluster center and the number of WATs decreases with the distance from the centre of a galaxy cluster. WATs usually settle in the cluster centre, i.e., at the bottom of the potential well \citep{Gu06}. The relative velocities of WATs with respect to the background are thought to be low \citep{Le84}. It is also known that the merger between two clusters could induce shocks which passage through the radio galaxies and could also affect the wide jet morphology \citep{Ga17}. NATs are usually field galaxies moving through their host cluster with high velocity $\sim 2500$ km~s$^{-1}$ (e.g., NGC 1265 \citet{Su05}). NATs and WATs originate from the same physical processes, but jet physics are different as a result of the different environments.

\subsection{Cluster mass and tailed radio galaxies}
The bending of jets is affected by the mass of the parent cluster in two ways. Firstly, the cluster mass is correlated with the density of the inter-cluster medium for which heavier clusters experience the denser inter-cluster medium.
Secondly, the mass of the cluster is correlated with the velocity dispersion of galaxies, which results in faster movement of galaxies in high mass clusters \citep{Mu15}. According to a simulation by \citet{Mu15}, there are no tailed radio galaxies in clusters with masses less than $10^{13}M_{\odot}$ and only a few in clusters with masses less than $10^{13.5}M\odot$. Another simulation suggests the presence of over 50000 galaxy clusters in the $10^{13}M_{\odot}-10^{14}M_{\odot}$ mass range and over 4000 galaxy clusters with masses larger than $10^{14}M_{\odot}$ which give proper favourable conditions for radio jet bending \citep{Go06}. The majority of tailed radio galaxies are seen in clusters with the mass range $10^{14}M_{\odot}$ and $10^{15}M_{\odot}$. Clusters above $10^{14.5}M_{\odot}$ are expected to host at least one-tailed galaxy at some point during their lifetime \citep{Mu15}. In tables 3 and 4, we summarise different properties of clusters corresponding to WAT and NAT galaxies detected in the current paper.
The mass and radius are two important properties of a cluster. The widely used parameters of a cluster are $r_{500}$, the radius within which the mean density of a cluster is 500 times of the critical density of the Universe ($\rho_{c}=\frac{3H^{2}(z)}{8\pi G}$), and $M_{500}$, the cluster mass within $r_{500}$. We calculate the cluster optical mass $M_{500}$ (in unit of $10^{14}M_{\odot}$) from cluster richness $(R_{L})$ which is closely correlated as \citep{We15}

\begin{equation}
 \log M_{500} = (1.08 \pm 0.02) \log R_{L} - (1.37 \pm0.02), 
\end{equation}

We find host clusters for {\bf 118} tailed sources using 22 known cluster catalogues summary of which is shown in Table \ref{tab:WAT-NAT-cat}. The maximum  clusters can be also found from the WHL cluster catalogue which provides optical richness and optical mass ($M_{500}$). The range of optical mass ($M_{500}$) of clusters lies in between $0.27\times10^{14}M_{\odot}-13.06\times10^{14}M_{\odot}$. Four of these clusters come from X-ray cluster catalogues. We calculate the corresponding optical masses of galaxy clusters from MCXC cluster catalogue using their optical properties from \citet{We12, We15}. We find that the majority (166 of 284 total) of tailed radio galaxies in our sample have no association with known clusters or groups. The detected WAT and NAT sources should be used to search for nearby clusters. There are 10 clusters found in relatively low-mass clusters (of the order of $10^{14}M_{\odot}$). 

\section{Summary}
\label{sec:sum}
We have presented 264 new tailed galaxy candidates from the visual inspection of the 150 MHz TGSS survey out of which $203$ are WAT type and $61$ are NAT type sources. We found $192$ sources from the northern sky and 72 sources from the southern sky. Because of the low radio frequency used in the survey, a significant number of steep-spectrum sources are seen.
The summary of the paper is as follows:

\begin{itemize}
\item We report a sample of 264 tailed radio galaxies. This makes it the largest sample of tailed radio galaxies discovered to date. Out of 264 new tailed galaxies, 203 are WAT type galaxies and 61 are NAT type galaxies.

\item We report a sample of 264 tailed radio galaxies. This makes it the largest sample of tailed radio galaxies discovered to date. Out of 264 new tailed galaxies, 203 are WAT type galaxies and 61 are NAT type galaxies.

\item The optical/IR counterparts of $261$ sources are found, and redshift is available for $112$ sources. About 45 per cent (92/203) WATs and 32 per cent (20/61) NATs are identified by redshift.

\item The total integrated flux density of sources in our sample at 150 MHz ranges from 0.1 Jy to as large as 20.1 Jy.

\item In our sample of tailed radio galaxies, two-tailed galaxies (WAT-J0932+5533 and NAT-J2038--2011) is hosted by quasars.

\item The high-resolution deep radio observations of TGSS-ADR images have enabled us to find tailed radio galaxies with less luminosity at 150 MHz. J0856+4829 is the least luminous WAT in our sample with $L_{150}=1.5\times10^{24}$ W Hz$^{-1}$ ($z=0.12$) and J0549--2520 is the least luminous NAT in our sample with $L_{150}=2.7\times10^{24}$ W Hz$^{-1}$ ($z=0.04$). J0225+4031 is the most luminous WAT in our sample with $L_{150}=1.9 \times10^{27}$ W Hz$^{-1}$ ($z\sim0.65$) and J1314+6220 is the most luminous NAT in our sample with $L_{150}=162.8\times10^{25}$ W Hz$^{-1}$ ($z=0.14$).

\item The spectral index peaks near $0.75$ for both WATs and NATs. For WAT, the total span of $\alpha_{150}^{1400}$ is from $0.02$ to $1.70$ and for NATs, the total span of $\alpha^{1400}_{150}$ is from $0.38$ to $2.04$.

\item The measured span of optical mass of host galaxy clusters for $M_{500}$ is $12.33\times10^{14}M_{\odot}-0.52\times10^{14}M_{\odot}$.

\item The range of the calculated density of galaxy clusters is from 206.7 (Mpc$^{-3}$) to 36.3 (Mpc$^{-3}$).

\end{itemize}
\section*{Acknowledgements}
We thank the whole TGSS team and the staff of the GMRT behind this survey. GMRT is run by the National Centre for Radio Astrophysics of the Tata Institute of Fundamental Research. This research has made use of the ``Aladin sky atlas" developed at CDS, Strasbourg Observatory, France and the NASA/IPAC Extragalactic Database (NED) operated by the Jet Propulsion Laboratory, California Institute of Technology. This publication makes use of data products from the Two Micron All Sky Survey, which is a joint project of the University of Massachusetts and the Infrared Processing and Analysis Center/California Institute of Technology, funded by the National Aeronautics and Space Administration and the National Science Foundation.
\section*{Data Availability Statement}
The data that support the plots within this paper and other findings of this study are available
from the corresponding author upon reasonable request. The TGSS ADR 1 images are available at \href{http://tgssadr.strw.leidenuniv.nl/doku.php}{http://tgssadr.strw.leidenuniv.nl/doku.php}.
\input table-final.tex

\bsp	
\label{lastpage}
\end{document}

%% file: table-final.tex
\begin{table*}
\caption{\bf Candidate WAT radio sources}
\label{tab:WAT}
  
\begin{threeparttable}
\begin{tabular}{cccccccccccll}
\hline
Cat&  Name     &R.A-core  &Dec-core &Separation  &$\rho$    &$F_{150}$ &$F_{1400}$ & ${\alpha}^{1.4}_{0.15}$ & $z$ &$L_{150}$ & Other	\\
N0.&           &(J2000.0) &(J2000.0)&(arsec)     &       &(Jy)      &(Jy)       &($\pm0.05$)              &     & W$Hz^{-1}$& Catalog\\
   &           &          &         &            &       &          &           &                         &     &$\times 10^{25}$     \\
(1)& (2)       &(3)       &(4)      & (5)        &(6)    &(7)       &(8)        &(9)                     &(10) &(11)            &(12)  \\

\hline
1  &J0001+5458 &00 01 21.8&+54 58 10&5.20  &0.90    &0.59      &0.10      &0.79  &--   &--            &1, 8\\%
2  &J0003--3556&00 03 11.5&-35 56 38&2.12  &0.99    &7.16      &--        &--    &0.05 &--            &14\\
3  &J0005+5359 &00 05 46.3&+53 59 57&3.50  &0.99   &0.67      &0.09      &0.89  &--   &--            &1, 8\\
4  &J0008--3347&00 08 49.6&-33 47 51&6.03  &0.89   &0.33      &0.03      &1.07  &--   &--            &1, 14\\
5  &J0022+2317 &00 22 24.4&+23 17 25&5.14  &0.99   &2.14      &0.41      &0.73  &0.13 &~~9.36        &1\\
6  &J0041+2104 &00 41 43.1&+21 04 26&5.01  &0.99   &0.35      &0.33      &0.02  &--   &--            &1\\
7  &J0050+0514 &00 50 44.6&+05 15 01&3.12  &0.99   &0.27      &0.10      &0.44  &0.27$^{b}$&~~5.42   &1\\%
8  &J0054+3339 &00 54 43.8&+33 39 01&6.58  &0.99   &0.33      &0.06      &0.76  &--   &--            &1, 4\\%
9  &J0104+8210 &01 04 25.1&+82 10 31&6.47  &0.89   &0.66      &0.19      &0.55  &--   &--            &1, 4\\
10 &J0106+4909 &01 06 48.6&+49 08 56&9.28  &0.99   &0.43      &0.07      &0.81  &--   &--            &1\\
11 &J0114+0029 &01 14 25.6&+00 29 33&6.69  &0.99   &0.36      &0.09      &0.62  &0.35 &13.46         &11\\%
12 &J0119+5838 &01 19 35.3&+58 38 54&4.71  &0.99   &0.55      &0.08      &0.86  &--   &--            &1, 8\\%
13 &J0120+1451 &01 20 01.1&+14 51 38&6.29  &0.88   &0.59      &0.16      &0.58  &0.05$^{b}$ &~~0.34  &1, 8\\
14 &J0123+3315 &01 23 39.9&+33 15 22&0.29  &0.99   &1.43      &0.10       &1.19   &0.02 &~~0.12        &1\\
15 &J0128+3448 &01 28 59.1&+34 48 42&6.18  &0.99   &0.73      &0.14       &0.73   &0.15 &04.32         &1,4\\
16 &J0139+7444 &01 39 25.2&+74 44 58&2.31  &0.99   &0.15      &0.04       &0.59   &--   &--            &1, 8\\%
17 &J0141+4506 &01 41 37.0&+45 06 07&2.10  &0.99   &0.31      &0.09       &0.55   &--   &--            &1\\%
18 &J0149+1403 &01 49 11.9&+14 03 01&6.22  &0.99   &0.46      &0.13       &0.56   &0.07 &~~0.53        &1\\%
19 &J0200+3935 &02 00 53.0&+39 35 01&6.46  &0.99   &1.80      &0.61       &0.48   &--   &--             &10\\%
20 &J0203+6702 &02 03 08.5&+67 02 52&1.95  &0.99   &0.89      &0.12       &0.89   &--   &--             &1, 4\\%
21 &J0204+0415 &02 04 28.9&+04 15 17&4.95  &0.99   &0.33      &0.10       &0.53   &0.14$^{a}$ &~~1.64   &11\\%
22 &J0205--4124&02 05 40.4&--41 24 13&6.08 &0.99   &1.10      &--         &--     &--   &--             &--  \\
23 &J0209+0950 &02 09 47.6&+09 50 02&3.68  &0.96   &4.69      &1.25       &0.59   &--   &--            &1\\
24 &J0225+4031 &02 25 44.9&+40 31 33&1.49  &0.95   &3.93      &0.91       &0.65   &0.14 &199.09         &3\\%
25 &J0228+3821 &02 28 30.3&+38 21 12&7.77  &0.99   &1.07      &0.22       &0.70   &0.03 &~~0.21         &4, 10\\%
26 &J0229+3942 &02 29 47.8&+39 42 29&1.34  &0.99   &0.98      &0.15       &0.84   &--   &--             &4, 10\\%
27 &J0245+8207 &02 45 20.2&+82 07 15&1.61  &0.99   &1.38      &0.18       &0.91   &--   &--            &1\\%
28 &J0257--0400&02 57 43.8&--03 59 59&3.61 &0.99   &1.10      &--         &--     &0.18$^{b}$ &--       &6, 11 \\%
29 &J0300+7438 &03 00 32.1&+74 38 59&--    &--     &2.36      &0.32       &0.89   &--   &--            &1, 2\\
30 &J0303+3856 &03 03 0.44&+38 56 39&9.98  &0.98   &0.95      &0.18       &0.74   &0.28$^{a}$ &~22.28   &1 ,4\\%
31 &J0306--1206&03 06 58.6&--12 06 43&3.31 &0.99   &5.94      &1.39       &0.65   &--   &--            &1, 2\\%
32 &J0310+4803 &03 10 16.9&+48 03 27&2.17  &0.99   &2.07      &0.55       &0.59   &--   &--            &1, 3\\%
33 &J0312+0644 &03 12 51.7&+06 44 48&7.42  &0.99   &0.41      &0.06       &0.86   &--   &--            &1\\%
34 &J0315+0507 &03 15 30.6&+05 07 46&5.30  &0.99   &0.38      &0.10       &0.59   &--   &--            &1\\%
35 &J0316+0922 &03 15 59.9&+09 23 01&3.89  &0.82   &0.45      &0.13       &0.55   &--   &--            &6\\
36 &J0317+4917 &03 17 33.3&+49 17 19&4.43  &0.89   &1.56      &0.40       &0.60   &--   &--            &1, 8\\%
37 &J0338+1544 &03 38 04.5&+15 44 19&11.7  &0.93   &0.16      &0.08       &0.31   &0.03$^{b}$ &~18.92  &1\\
38 &J0348+1924 &03 48 28.4&+19 24 19&10.1  &0.98   &1.23      &0.23       &0.75   &--   &--            &1\\%
39 &J0401+3853 &04 01 4.30&+38 53 00&4.15  &0.65   &0.55      &0.13       &0.64   &--   &--            &1, 10\\%
40 &J0402+1929 &04 02 13.0&+19 29 07&8.54  &0.64   &2.58      &0.78       &0.53   &--   &--            &1\\
41 &J0405--2610&04 05 15.9&--26 10 58&1.23 &0.99   &0.93      &0.18       &0.73   &--   &--            &1  \\
42 &J0415+3428 &04 15 13.5&+34 28 29&3.23  &0.99   &0.68      &0.14       &0.70   &--   &--            &1, 4, 8\\
43 &J0436+0603 &04 36 22.5&+06 03 12&7.63  &0.99   &1.45      &0.24       &0.70   &--   &--            &1\\ 
44 &J0458+8206 &04 58 38.3&+82 06 35&4.50  &0.99   &0.51      &0.08       &0.82   &--   &--            &2, 4\\%
45 &J0505+1056 &05 05 08.1&+10 56 08&2.11  &0.99   &1.34      &0.28       &0.70   &--   &--            &1\\
46 &J0508+2754 &05 08 12.3&+27 54 15&4.64  &0.87   &2.36      &0.40       &0.79   &--   &--            &1\\
47 &J0520+1356 &05 20 12.5&+13 56 01&4.77  &0.99   &0.95      &0.22       &0.65   &--   &--            &1, 3\\
48 &J0522+2807 &05 22 46.0&+28 06 40&6.65  &0.99   &1.91      &--         &--     &--   &--            &--\\%
49 &J0528+5601 &05 28 38.3&+56 01 49&1.33  &0.99   &0.43      &0.07       &0.81   &--   &--            &4\\
50 &J0533+4727 &05 33 54.8&+47 27 07&10.5  &0.82   &2.49      &0.34       &1.12   &--   &--            &1, 4, 10\\%
51 &J0542+7902 &05 42 58.0&+79 02 31&--    &--     &0.61      &0.16       &0.59   &--   &--            &15\\ 
52 &J0602+1937 &06 02 37.2&+19 37 05&4.57  &0.99    &0.85      &0.14       &0.80   &--  &--             &1, 8\\%
53 &J0602+2911 &06 02 34.1&+29 11 06&9.66  &0.82    &2.99      &0.48       &0.81   &--  &--             &1, 2\\ 
54 &J0608+8058 &06 08 32.1&+80 58 12&1.52  &0.99    &0.77      &0.16       &0.70   &--  &--             &1, 4\\
55 &J0623+0504 &06 23 11.5&+05 04 12&4.29  &0.99    &0.85      &0.15       &0.77   &--  &--             &1, 6\\
56 &J0626+2838 &06 26 38.6&+28 38 42&2.16  &0.99    &0.72      &0.09       &0.93   &--  &--             &1, 8, 9\\
57 &J0630+2306 &06 30 31.4&+23 06 05&6.02  &0.99    &0.52      &0.09       &0.78   &--  &--             &16\\
58 &J0631+2500 &06 31 25.8 &+25 00 48&1.94 &0.99  &13.3      &1.08       &1.12   &0.08$^{a}$&~~3.39   &1\\
59 &J0640+1020 &06 40 41.6 &+10 20 36&6.20 &0.95  &0.56      &0.12       &0.68   &--  &--             &1,15\\
60 &J0645+1340 &06 45 28.0 &+13 40 25&--   &--    &1.01      &0.13       &0.91   &--  &--             &1, 2\\
\hline
\end{tabular}
\end{threeparttable}
\end{table*}

\begin{table*}
\contcaption{\bf Candidate WAT radio sources}
\begin{threeparttable}
\begin{tabular}{ccccccccccccll}
\hline
Cat&  Name     &R.A-core  &Dec-core &Separation  &$\rho$    &$F_{150}$ &$F_{1400}$ & ${\alpha}^{1.4}_{0.15}$ & $z$ &$L_{150}$ & Other	\\
N0.&           &(J2000.0) &(J2000.0)&(arsec)     &       &(Jy)      &(Jy)       &($\pm0.05$)              &     & W$Hz^{-1}$& Catalog\\
   &           &          &         &            &       &          &           &                         &     &$\times 10^{25}$     \\
(1)& (2)       &(3)       &(4)      & (5)        &(6)    &(7)       &(8)        &(9)                     &(10) &(11)            &(12)  \\
\hline
61 &J0653+6919 &06 53 17.8 &+69 19 39&4.41   &0.99    &8.18      &1.38       &0.79   &--   &--            &1, 2, 4\\
62 &J0655+0412 &06 55 20.3 &+04 12 09&3.14   &0.99    &2.49      &0.30       &0.94   &--  &--             &6\\
63 &J0700+2736 &07 00 01.0 &+27 36 28&5.94   &0.99    &1.73      &0.31       &0.76   &--  &--             &1,9\\
64 &J0708+2928 &07 08 48.5 &+29 28 13&5.24   &0.99    &1.22      &0.16       &0.90   &0.19&~12.44         &1,9\\%
65 &J0714+1334 &07 14 10.5 &+13 34 49&6.33   &0.99    &0.45      &0.10       &0.67   &0.54&~46.12        &1, 8\\ %
66 &J0715--3044&07 15 24.8 &--30 43 45&10.1  &0.96    &1.65      &0.30       &0.76   &--  &--            &1  \\%
67 &J0724+5010 &07 24 31.5 &+50 10 53&3.96   &0.61    &1.00      &0.15       &0.84   &--  &--            &1, 8, 11\\%
68 &J0736+2412 &07 36 21.8 &+24 12 14&5.29   &0.99    &4.88      &0.44       &1.07   &0.15&~30.35        &1, 2\\
69 &J0752+0814 &07 52 57.0 &+08 14 33&7.20   &0.93    &0.64      &0.04       &1.24   &--  &--            &1, 2\\%
70 &J0756+0107 &07 56 7.31 &+01 07 48&4.37   &0.76    &1.30      &0.41       &0.51   &--  &--            &1\\
71 &J0757+3640 &07 57 53.2 &+36 40 22&6.91   &0.99    &1.44      &0.32       &0.67   &0.12$^{a}$&~~5.25  &12, 13\\%
72 &J0805+1614 &08 05 44.8 &+16 14 05&2.46   &0.99    &1.01      &0.18       &0.77   &0.10$^{b}$&~~2.48  &12\\%
73 &J0818+5437 &08 18 06.6 &+54 37 32&3.18   &0.99    &3.31      &0.51       &0.83   &0.12$^{b}$&~12.28  &11\\%
74 &J0831+6104 &08 31 48.8 &+61 04 59&9.36   &0.98    &1.24      &0.14       &0.97   &--  &--            &1\\%
75 &J0841+4451 &08 41 04.3 &+44 51 04&1.86   &0.65    &0.48      &0.07       &0.86   &0.27&~10.67         &1, 10, 11\\%
76 &J0856+4829 &08 56 1.17 &+48 29 20&10.2   &0.99    &0.44      &0.16       &0.45   &0.12$^{b}$&~~0.15    &4, 8, 11\\%
77 &J0912+1600 &09 12 35.2 &+16 00 01&7.33   &0.99    &0.10      &0.02       &0.72   &0.08$^{b}$&~~1.55    &1\\
78 &J0917+5509 &09 17 08.0 &+55 09 08&4.34   &0.99    &1.30      &0.22       &0.79   &0.19 &13.13            &4,8\\
79 &J0932+5533 &09 32 01.0 &+55 33 47&0.80   &0.99    &0.81      &0.17       &0.69   &0.26$^{b}$&~15.92   &1, 11\\%
80 &J0940+1131 &09 40 34.4 &+11 31 40&3.94   &0.99    &2.76      &0.58       &0.69   &0.08$^{b}$&~~1.32        &1\\%
81 &J0944+0247 &09 44 43.4 &+02 47 57&2.96   &0.68    &3.73      &0.35       &1.05   &0.22 &03.64            &1,6\\
82 &J0951--0757&09 51 30.2 &--07 57 51&3.81  &0.99    &0.72      &--         &       &--  &--             &-- \\%
83 &J0957--0644&09 57 00.4 &--06 44 26&8.44  &0.99    &2.27      &0.40       &0.77   &0.13&~~9.96         &1,6 \\%
84 &J1003+1019 &10 03 41.9 &+10 20 01&5.33   &0.99    &0.69      &0.20       &0.55   &0.18&~~0.89          &1, 11, 16\\%
85 &J1008--4055&10 08 11.0 &--40 55 44&9.12  &0.99    &0.66      &--         &--     &--  &--              &14 \\%
86 &J1009--1504&10 09 07.3 &--15 04 04&1.67  &0.92    &1.96      &0.12       &1.25   &--   &--            &1, 2 \\%
87 &J1012+0841 &10 12 06.4 &+08 41 33&5.88   &0.99    &0.87      &0.21       &0.63   &0.09 &01.72            &1,11\\%
88 &J1015+1221 &10 15 41.1 &+12 21 01&7.41   &0.99    &0.35      &0.10       &0.56   &--- &--             &1, 11\\%
89 &J1019+7020 &10 19 56.1 &+70 20 22&5.42   &0.99    &0.72      &0.12       &0.80   &--- &--             &1, 4, 8\\
90 &J1022+5006 &10 22 28.4 &+50 06 20&4.24   &0.65    &0.73      &0.03       &1.42   &0.16 &05.51            &2,4 \\
91 &J1032+3151 &10 32 18.0 &+31 51 42&3.60   &0.99    &0.79      &0.16       &0.71   &0.35&~30.40        &1, 9, 11\\
92 &J1034+0736 &10 34 10.0 &+07 36 05&3.76   &0.97    &0.92      &0.12       &0.91   &0.58$^{b}$&123.92   &1, 2, 11\\%
93 &J1038--3346&10 38 47.9 &--33 46 35&3.97  &0.99    &0.50      &0.06       &0.94   &--   &--      &1 \\
94 &J1042+0237 &10 42 31.3 &+02 37 10&8.65   &0.99    &0.87      &0.12       &0.88   &0.18$^{b}$&~~7.85   &1, 6, 11\\ 
95 &J1046--2911&10 46 09.9 &--29 21 10&3.48  &0.99    &2.14      &0.55       &0.60   &0.06&~~1.81       &1 \\
96 &J1048+3532 &10 48 49.0 &+35 32 01&3.51   &0.99    &1.94      &0.22       &0.97   &0.39&104.11       &1, 4, 13\\
97 &J1050+0432 &10 50 57.6 &+04 32 17&10.2   &0.99    &0.34      &0.08       &0.64   &0.12&~~1.23       &1\\%
98 &J1050--2405&10 50 34.5 &--24 05 57&5.95  &0.99    &3.19      &0.67       &0.69   &0.03&~~0.63       &1 \\
99 &J1051+1825 &10 51 57.7 &+18 25 38&5.61   &0.99    &0.49      &0.11       &0.66   &0.55$^{b}$&~~1.48 &1, 11\\%
100&J1056+0255 &10 56 17.0 &+02 55 26&1.94   &0.97    &0.54      &0.07       &0.91   &0.39 &28.40            &1\\
101&J1058+0136 &10 58 07.5 &+01 36 19&2.34   &0.99    &1.26      &0.04       &1.54   &0.04&~~0.47        &1, 2\\
102&J1106--4018&11 06 58.0 &--40 18 55&10.5  &0.93    &0.65      &--         &--     &--  &--            &14 \\
103&J1108--4424&11 08 11.0 &--44 24 30&1.68  &0.62    &0.55      &--         &--     &--  &--            &14 \\%
104&J1108+2610 &11 08 12.5 &+26 10 34&5.96   &0.95    &0.66      &0.12       &0.76   &0.17$^{b}$    &~~5.16 &11, 13\\%
105&J1116--3010&11 16 00.1 &--30 10 08&1.95  &0.99    &0.73      &0.12       &0.80   &--      &--            &1 \\%
106&J1118+2754 &11 18 59.4 &+27 54 07&7.10   &0.99    &2.05      &0.52       &0.61   &0.06    &~~1.74        &11\\
107&J1119+6317 &11 19 33.2 &+63 17 17&4.39   &0.99    &0.42      &0.11       &0.59   &0.16 &02.80            &1,4,11\\
108&J1120+2912 &11 20 38.5 &+29 12 34&5.26   &0.99    &0.28      &0.08       &0.56   &0.24$^{b}$    &~~4.48  &1, 11\\%
109&J1130+2524 &11 30 48.8 &+25 24 36&2.99   &0.99    &0.41      &0.09       &0.67   &0.14$^{b}$    &~~2.08  &1, 11\\%
110&J1131+4408 &11 31 10.1 &+44 08 15&10.1   &0.85    &0.49      &0.06       &0.94   &--      &--            &1, 2, 10\\%
111&J1132+6311 &11 32 51.0 &+63 11 44&3.59   &0.99    &1.90      &0.43       &0.66   &0.11$^{b}$    &~~1.30 &1, 4\\%
112&J1141--3357&11 41 28.9 &--35 57 10&5.45  &0.99    &0.32      &--         &--     &--      &--           &1 \\%
113&J1142+1102 &11 42 54.3 &+11 01 33&8.53   &0.99    &1.30      &0.34       &0.60   &0.15$^{b}$    &~~7.57 &17\\
114&J1151+0422 &11 51 46.9 &+04 22 23&2.73   &0.86    &0.39      &0.09       &0.65   &0.13 &01.68            &1,11\\
115&J1155+5755 &11 55 58.3 &+57 55 27&9.30   &0.93    &0.38      &0.11       &0.55   &--      &--           &2, 11\\
116&J1156+3432 &11 56 7.40 &+34 32 47&5.32   &0.98    &1.03      &0.15       &0.86   &--      &--          &1, 11, 13\\
117&J1158+2117 &11 58 37.2 &+21 17 11&5.98   &0.99    &0.36      &0.07       &0.73   &--   &--               &1,11\\%
\hline
\end{tabular}
\end{threeparttable}
\end{table*}

\begin{table*}
\contcaption{\bf Candidate WAT radio sources}
\begin{threeparttable}
\begin{tabular}{ccccccccccccll}
\hline

Cat&  Name     &R.A-core  &Dec-core &Separation  &$\rho$   &$F_{150}$ &$F_{1400}$ & ${\alpha}^{1.4}_{0.15}$ & $z$ &$L_{150}$ & Other	\\
N0.&           &(J2000.0) &(J2000.0)&(arsec)     &       &(Jy)      &(Jy)       &($\pm0.05$)              &     & W$Hz^{-1}$& Catalog\\
   &           &          &         &            &       &          &           &                         &     &$\times 10^{25}$     \\
(1)& (2)       &(3)       &(4)      & (5)        &(6)    &(7)       &(8)        &(9)                     &(10) &(11)            &(12)  \\

\hline
118&J1200+2942 &12 00 46.1&+29 42 58&7.52  &0.99     &1.20      &0.24       &0.72   &0.16    &~~8.18       &11, 13\\
119&J1202+5802 &12 02 3.80&+58 02 08&3.59  &0.99     &5.70      &0.85       &0.85   &0.10$^{b}$    &~14.42 &1, 11\\%
120&J1205+3204 &12 05 14.5&+32 04 17&5.03  &0.99     &0.72      &0.18       &0.62   &0.16 &04.83            &11,13\\
121&J1206+3152 &12 06 47.6&+31 52 31&1.59  &0.89     &0.80      &0.17       &0.69   &0.25 &14.48            &11,13\\%
122&J1212--4545&12 12 20.7&--45 45 20&1.99 &0.99     &1.96      &--         &--     &0.16    &--           &6 \\%
123&J1221--4046&12 21 24.4&--40 46 15&0.96 &0.98     &0.32      &--         &--     &--      &--           &14 \\%
124&J1234--1045&12 34 29.6&--10 45 25&7.78 &0.95     &0.34      &0.06       &0.77   &--   &--            &1\\%
125&J1242--3613&12 42 01.0&--36 13 40&2.28 &0.68     &1.52      &0.28       &0.75   &--   &--            &1\\%
126&J1242+5021 &12 42 7.40&+50 21 47&1.50  &0.99     &0.34      &0.10       &0.54   &0.15 &01.96            &1,8\\%
127&J1249+0144 &12 49 42.9&+01 44 18&3.79  &0.94     &0.50      &0.20       &0.41   &--      &--          &1\\%
128&J1304+1041 &13 04 17.9&+10 40 35&2.53  &0.99     &0.44      &0.05       &0.97   &0.11$^{b}$    &~~1.12 &1\\%
129&J1304+6439 &13 04 28.8&+64 39 38&4.48  &0.88     &1.83      &0.31       &0.79   &0.23$^{b}$    &~27.98 &1, 4, 8\\%
130&J1307+5651 &13 07 43.7&+56 51 03&2.23  &0.99     &0.72      &0.13       &0.76   &0.24 &08.78            &4,8,11\\%
131&J1311--0120&13 11 31.7&--01 20 00&4.88 &0.99     &1.09      &0.05       &1.37   &0.18    &~10.68        &1\\%
132&J1315+4841 &13 15 30.5&+48 41 14&7.67  &0.86     &0.45      &0.07       &0.83   &0.68$^{b}$    &~85.27  &1\\%
133&J1325+5736 &13 25 11.2&+57 36 01&3.98  &0.99     &1.63      &0.10       &1.24   &0.12$^{b}$    &~~5.24  &1\\%
134&J1331--0544&13 31 24.3&--05 44 36&4.05 &0.99     &2.92      &0.67       &0.66   &0.15    &~~5.88        &1\\
135&J1412+7420 &14 12 25.8&+74 20 19&1.40  &0.99     &1.23      &0.21       &0.79   &0.21    &~13.80        &1\\%
136&J1414+0143 &14 14 32.6&+01 43 54&6.12  &0.99     &0.56      &0.19       &0.48   &0.05$^{b}$    &~~0.32  &1\\%
137&J1416+0219 &14 16 13.4&+02 19 08&6.46  &0.99     &0.70      &0.25       &0.46   &0.16$^{b}$    &~~3.94  &1\\%
138&J1440+0328 &14 40 39.0&+03 28 37&3.94  &0.99     &0.61      &--         &--     &0.03    &--            &2\\%
139&J1449+3959 &14 49 0.86&+40 00 44&2.09  &0.99     &1.68      &0.26       &0.83   &---     &--            &1, 11, 13\\%
140&J1450+4418 &14 50 39.8&+44 18 29&3.10  &0.99     &1.45      &0.28       &0.73   &0.29$^{b}$   &~33.94   &1, 13\\%
141&J1501+0752 &15 01 57.4&+07 52 27&10.2  &0.97     &2.55      &0.50       &0.72   &0.66$^{b}$   &362.4 &1, 11\\%
142&J1508+3554 &15 08 34.1&+35 54 28&9.36  &0.68     &0.90      &0.18       &0.72   &0.12$^{b}$    &~~3.68    &4, 9\\%
143&J1509+3327 &15 09 59.7&+33 27 46&9.84  &0.99     &0.85      &0.17       &0.72   &0.12$^{b}$    &~~3.12    &1, 13\\%
144&J1543--4345&15 43 13.5&--43 45 01&5.55 &0.99     &3.39      &--         &--     &--      &--            &-- \\
145&J1553+1530 &15 53 45.5&+15 30 14&3.00  &0.98     &0.72      &0.14       &0.73   &0.13$^{b}$    &~3.14   &1, 8, 11\\%
146&J1604+2355 &16 04 56.7&+23 55 58&3.13  &0.99     &4.13      &0.66       &0.82   &0.03    &~~2.70        &1, 13\\%
147&J1612+2929 &16 12 35.4&+29 29 05&3.85  &0.99     &0.55      &0.11       &0.72   &0.03    &~~1.78        &1, 13\\%
148&J1615+4711 &16 15 44.1&+47 11 45&9.56  &0.99     &4.88      &0.57       &0.96   &--      &--            &1, 4, 11\\
149&J1616+0926 &16 16 53.2&+09 26 36&7.84  &0.89     &4.88      &0.23       &1.36   &0.20 &60.75            &1\\%
150&J1620+2521 &16 20 35.8&+25 21 04&1.23  &0.99       &0.62      &0.18       &0.55   &---     &--            &1, 13\\
151&J1627--4336&16 27 01.0&--43 36 20&7.82 &0.99     &2.35      &--         &--     &--      &--            &6 \\%
152&J1631+0115 &16 31 30.4&+01 16 14&2.81  &0.95     &0.43      &0.06       &0.88   &---     &--            &1\\%
153&J1631+0501 &16 31 06.1&+05 01 27&8.74  &0.84     &1.19      &0.17       &0.86   &---     &--            &1\\%
154&J1636+2718 &16 36 5.20&+27 18 35&4.28  &0.99     &1.78      &0.43       &0.63   &0.13$^{b}$&~~7.76      &1, 9, 11\\%
155&J1658+6256 &16 58 47.1&+62 56 25&3.79  &0.99     &1.31      &0.29       &0.67   &0.10$^{b}$&~~3.24      &1, 4, 8\\%
156&J1701+6413 &17 01 32.0&+64 13 38&6.11  &0.99     &0.37      &0.07       &0.74   &---     &--            &1, 4, 8\\%
157&J1711+0449 &17 11 53.0&+04 49 28&8.22  &0.99     &1.37      &0.16       &0.96   &---     &--            &1, 8\\
158&J1711+1351 &17 11 50.9&+13 51 51&5.82  &0.99     &0.44      &0.24       &0.27   &---     &--            &1\\
159&J1717+3734 &17 17 25.4&+37 34 59&5.14  &0.99     &0.37      &0.15       &0.40   &0.09    &00.71         &1,13\\
160&J1719+1557 &17 19 39.6&+15 57 40&1.08  &0.99     &3.60      &0.08       &1.70   &---     &--            &1\\
161&J1727--2815&17 27 01.1&--28 15 30&1.04 &0.99     &2.34      &0.06       &1.64   &--      &--            &1, 15 \\
162&J1729--4330&17 29 56.8&--43 30 55&2.61 &0.99     &0.76      &0.19       &0.62   &--      &--            &1, 6 \\
163&J1735+3137 &17 35 06.5&+31 37 51&9.00  &0.99     &0.56      &0.12       &0.69   &0.28    &~~2.03        &1, 9\\
164&J1736+1414 &17 36 19.4&+14 14 58&1.59  &0.52     &0.46      &0.07       &0.84   &--      &--            &1\\%
165&J1741--3831&17 41 01.0&--38 32 05&3.65 &0.99     &0.76      &0.19       &0.62   &--      &--            &1, 6 \\
166&J1741+1720 &17 41 39.2&+17 20 34&9.43  &0.99    &9.05      &1.58       &0.78   &0.06    &~~7.78        &1\\%
167&J1748+1400 &17 48 26.2&+14 00 54&7.16  &0.82    &1.14      &0.11       &1.04   &--      &--            &1, 2\\%
168&J1752+5218 &17 52 52.6&+52 17 56&6.35  &0.99    &0.19     &0.03       &0.82   &--      &--            &1\\%
169&J1801+2109 &18 01 48.5&+21 09 27&7.23  &0.86    &4.48      &0.84       &0.75   &--      &--            &1\\%
170&J1814--2302&18 14 39.5&--23 02 38&10.5 &0.99    &1.73      &0.09       &1.32   &--      &--            &1 \\%
171&J1816+1415 &18 16 29.2&+14 15 09&8.54  &0.99    &3.28      &0.60       &0.76   &--      &--            &1, 2\\%
172&J1818+1739 &18 18 08.5&+17 39 42&10.5  &0.98    &0.99      &0.10       &1.02   &--      &--            &1\\%
173&J1820+0855 &18 20 2.20&+08 55 14&8.40  &0.72    &1.87      &0.25       &0.90   &--      &--             &15, 16\\%
174&J1820+2011 &18 20 25.5&+20 11 55&0.53  &0.99    &0.37      &0.07       &0.74   &--      &--             &1\\%
175&J1826+3449 &18 26 50.2&+34 49 40&2.54  &0.93    &3.72      &0.12       &1.53   &--      &--             &1, 9\\%
176&J1827+6832 &18 27 31.8&+68 32 15&3.47  &0.99    &0.76      &0.15       &0.72   &--      &--            &1, 5\\%
177&J1856+3556 &18 56 24.3&+35 56 18&5.96  &0.98    &1.87      &0.57       &0.53   &--      &--             &1, 2, 9\\%

\hline
\end{tabular}
\end{threeparttable}
\end{table*}

\begin{table*}
\contcaption{\bf Candidate WAT radio sources}
\begin{threeparttable}
\begin{tabular}{ccccccccccccll}
\hline

Cat&  Name     &R.A-core  &Dec-core &Separation  &$\rho$    &$F_{150}$ &$F_{1400}$ & ${\alpha}^{1.4}_{0.15}$ & $z$ &$L_{150}$ & Other	\\
N0.&           &(J2000.0) &(J2000.0)&(arsec)     &       &(Jy)      &(Jy)       &($\pm0.05$)              &     & W$Hz^{-1}$& Catalog\\
   &           &          &         &            &       &          &           &                         &     &$\times 10^{25}$     \\
(1)& (2)       &(3)       &(4)      & (5)        &(6)    &(7)       &(8)        &(9)                     &(10) &(11)            &(12)  \\
\hline

178&J1906--0033 &19 06 16.0&--00 33 29&3.82 &0.99    &0.92      &0.18       &0.73   &--      &--             &1, 8\\%
179&J1922+3920  &19 22 14.7&+39 20 19&4.76  &0.99    &2.36      &0.58       &0.62   &--      &--             &1, 9\\%
180&J1926+4124  &19 26 47.6&+41 24 48&7.13  &0.99    &0.62      &0.22       &0.46   &--      &--             &1\\%
181&J1930--1509 &19 30 01.9&--15 09 19&7.25 &0.99    &10.01     &1.48       &0.85   &0.08    &~15.58        &1, 15\\%
182&J1944+7815  &19 44 11.2&+78 16 18&5.56  &0.99    &2.80      &0.56       &0.72   &--      &--            &1\\%
183&J1946+0201  &19 46 4.40&+02 01 54&4.26  &0.99    &3.06      &0.30       &1.03   &--      &--            &1\\%
184&J1950--2817 &19 50 45.9&--28 17 39&3.37 &0.99    &4.25      &0.62       &0.86   &--      &--            &1 \\%
185&J2000--0101 &20 00 52.7&--01 01 09&4.29 &0.99    &0.50      &0.12       &0.63   &--      &--            &1, 6\\%
186&J2015+1144  &20 15 20.9&+11 44 49&2.97  &0.99    &0.49      &0.13       &0.59   &--      &--            &1\\
187&J2042--0310 &20 42 56.5&--03 10 22&6.48 &0.99    &0.49      &0.16       &0.50   &0.14    &~~2.32        &1 \\
188&J2049+3526  &20 49 53.2&+35 26 52&8.08  &0.99    &0.44      &0.11       &0.62   &--      &--            &1, 9\\
189&J2102--0921 &21 02 15.7&--09 21 15&5.52 &0.99    &1.38      &0.42       &0.53   &0.08    &~~2.12        &1 \\%
190&J2108+6049  &21 08 41.1&+60 49 03&4.21  &0.99    &0.63      &0.17       &0.58   &--      &--            &1, 8\\
191&J2111+3335  &21 11 37.7&+33 35 12&8.49  &0.99    &0.63      &0.11       &0.78   &--      &--             &1, 2\\
192&J2138+8307  &21 38 49.8&+83 07 04&2.84  &0.99    &0.78      &0.31       &0.41   &0.14    &~~3.82         &1, 2\\
193&J2139--3218 &21 39 30.5&--32 18 38&5.00 &0.83    &0.94      &0.24       &0.61   &--      &--             &1, 6\\
194&J2144+8015  &21 44 0.71&+80 15 12&4.63  &0.99    &0.35      &0.12       &0.47   &--      &--             &1, 4\\
195&J2144--3105 &21 44 04.8&--31 05 18&1.71 &0.98    &0.49      &0.11       &0.66   &0.18    &~~4.27         &1, 4 \\%
196&J2157+0037  &21 57 31.4&+00 37 57&5.24  &0.99    &1.12      &0.24       &0.69   &0.39$^{b}$    &~54.80         &1, 6\\%
197&J2158+6014  &21 58 46.9&+60 14 41&5.03  &0.99    &0.51      &0.09       &0.77   &--      &--             &1, 8\\
198&J2212+1304  &22 12 54.4&+13 04 36&1.60  &0.99    &1.76      &0.07       &1.44   &0.15 &11.54            &1\\%
199&J2300+1426  &23 00 47.4&+14 26 03&4.48  &0.99    &1.29      &0.23       &0.77   &0.15$^{b}$ &~~8.46   &1\\%
200&J2301--4333 &23 01 17.7&--43 33 24&4.89 &0.99    &1.02      &--         &--     &--   &--            &6, 14 \\%
201&J2310+0734  &23 10 22.6&+07 34 53&9.33  &0.99    &0.62      &0.04       &1.22   &0.04    &~~0.03         &1\\%
202&J2322+4157  &23 22 43.3&+41 58 14&7.32  &0.99    &1.09      &0.39       &0.46   &0.11$^{a}$    &~3.82          &1\\%
203&J2348+0043  &23 48 23.9&+00 43 47&1.17  &0.99    &0.66      &0.09       &0.89   &0.36    &~28.63         &1, 2, 11\\

\hline
\end{tabular}
\begin{tablenotes}
      \small
     \item 1: NVSS \citep{Con98}; 2: VLSS \citep{Co07}; 3: 4C \citep{Pi65, Go67, Ca69};  4: 6C \citep{Ba85, Ha88, Ha90, Ha91, Ha93a, Ha93b}; 5: 7C \citep{Mc90, Ko94, Wa96, Ve98}; 6: PMN \citep{Gri94}; 7: PKS \citep{Bo64}; 8: 87GB \citep{Gr91}; 9: B2 \citep{Co70, Co72, Co73, Fr74}; 10: B3 \citep{Fi85}; 11: VFK \citep{Va15}; 12: CRATES \citep{He07}; 13: FIRST \citep{Be95}; 14: SUMSS \citep{Mau03}:; 15: 2MASS \citep{Skr06}; 16: MG2\citep{Be86}; 17: MRC \citep{La81} ; 18: SSTSL \citep{Ra12}
    \end{tablenotes}
\end{threeparttable}
\end{table*}

\begin{table*}
\caption{\bf Candidate NAT radio sources}
\label{tab:NAT}
\begin{threeparttable}
\begin{tabular}{ccccccccccccll}
\hline

Cat&  Name     &R.A-core  &Dec-core &Separation  &$\rho$    &$F_{150}$ &$F_{1400}$ & ${\alpha}^{1.4}_{0.15}$ & $z$ &$L_{150}$ & Other	\\
N0.&           &(J2000.0) &(J2000.0)&(arsec)     &       &(Jy)      &(Jy)       &($\pm0.05$)              &     & W$Hz^{-1}$& Catalog\\
   &           &          &         &            &       &          &           &                         &     &$\times 10^{25}$     \\
(1)& (2)       &(3)       &(4)      & (5)        &(6)    &(7)       &(8)        &(9)                     &(10) &(11)            &(12)  \\

\hline
1  &J0003+5745  &00 03 18.9&+57 45 28&7.33  &0.99      &0.43      &0.16       &0.44   &--   &--              &8\\%
2  &J0041--0922 &00 41 30.9&--09 22 33&7.23 &0.99      &10.5      &--         &--     &--   &--              &17\\
3  &J0041--4346 &00 41 29.5&--43 47 05&4.27 &0.99      &0.73      &--         &--     &--   &--              &18\\
4  &J0102--0050 &01 02 39.8&--00 50 41&2.32 &0.99      &0.43      &0.13       &0.53   &--   &--              &1, 11\\
5  &J0141+1623  &01 41 35.5&+16 23 36&4.87  &0.99      &1.05      &0.16       &0.84   &--   &--             &1, 8\\
6  &J0148--3155 &01 48 15.4&--31 55 20&3.23 &0.99      &0.73      &0.08       &0.97   &0.14 &~~9.42         &1  \\%
7  &J0223+4300  &02 23 19.4&+42 59 47&4.72  &0.99      &20.1      &5.56       &0.57   &0.02 &~17.88         &10\\
8  &J0228--2814 &02 28 26.0&--28 14 13&4.14 &0.99      &3.45      &0.68       &0.72   &0.20 &~17.38         &1  \\
9  &J0236+1202  &02 36 26.5&+12 02 51&1.14  &0.93      &0.49      &0.06       &0.94   &--   &--             &1\\
10 &J0303+6605  &03 03 0.44&+66 05 43&4.45  &0.99      &2.38      &0.33       &0.88   &--   &--             &1\\%
11 &J0303+1610  &03 03 57.7&+16 10 41&5.77  &0.99      &0.74      &0.29       &0.41   &--   &--             &1\\%
12 &J0335+6517  &03 35 9.20&+65 17 04&1.10  &0.81      &0.38      &0.20       &0.28   &--   &--             &1\\
13 &J0411+6133  &04 11 50.0&+61 33 10&1.72  &0.99      &1.18      &0.23       &0.73   &--   &--             &1\\
14 &J0516+3531  &05 16 19.7&+35 31 49&4.09  &0.99      &0.46      &0.12       &0.60   &--   &--             &1\\
15 &J0523+2427  &05 23 34.9&+24 27 27&10.1  &0.98      &0.96      &0.20       &0.70   &--   &--             &9\\
16 &J0541+2842  &05 41 13.4&+28 42 47&10.7  &0.91      &7.24      &1.40       &0.73   &--   &--             &1, 3\\
17 &J0549--2520 &05 49 21.6&--25 20 47&4.61 &0.99      &0.75      &0.20       &0.59   &0.04 &~~0.27         &1, 2  \\
18 &J0603+5108  &06 03 53.9&+51 08 29&7.33  &0.99      &1.28      &0.21       &0.80   &--   &--             &1\\
19 &J0603+5619  &06 03 28.0&+56 19 29&9.14  &0.99      &6.36      &0.01       &1.74   &--   &--             &1\\%
20 &J0704+6318  &07 04 28.7&+63 18 39&4.43  &0.99      &6.67      &0.03       &2.04   &0.09 &~14.87        &1, 2\\%
21 &J0709+5100  &07 09 53.2&+51 00 58&1.78  &0.99      &1.32      &0.16       &0.94   &--   &--             &1, 8, 11\\%
22 &J0728--0008 &07 28 50.3&--00 08 21&7.39 &0.99      &3.49      &0.78       &0.67   &--   &--             &--\\
23 &J0730+4051  &07 30 43.3&+40 51 52&1.83  &0.93      &0.93      &0.54       &0.24   &0.12$^{b}$ &~~3.22   &1, 11\\%
24 &J0735+2510  &07 35 42.3&+25 10 16&10.6  &0.96      &2.08      &0.23       &0.98   &--   &--             &1, 11\\
25 &J0802+6345  &08 02 02.5&+63 45 55&3.27  &0.99      &0.78      &0.07       &0.98   &0.07$^{a}$ &~~1.18   &4, 11, 14\\%
26 &J0810--4923 &08 10 42.6&--49 23 54&5.12 &0.99      &11.1      &--         &--     &--   &--             &--     \\
27 &J0815+3710  &08 15 45.6&+37 10 24&6.34  &0.96      &0.51      &0.13       &0.61   &--   &--            &4\\%
28 &J0818--0004 &08 19 0.50&--00 04 57&4.94 &0.99      &0.42      &0.06       &0.87   &--   &--            &1\\
29 &J0928+4854  &09 28 22.1&+48 54 33&6.00  &0.80      &0.64      &0.11       &0.78   &--   &--            &1, 4, 8\\%
30 &J0953+7057  &09 53 59.7&+70 57 34&2.93  &0.99      &4.88      &0.26       &1.31   &0.18$^{a}$ &~47.33  &1, 4, 8\\%
31 &J1108--4823 &11 08 51.0&--48 23 50&1.81 &0.99      &1.02      &--         &--     &--   &--             &14 \\
32 &J1111+4050  &11 11 39.7&+40 50 24&1.21  &0.79      &6.65      &0.98       &0.85   &0.07$^{b}$ &~~7.91   &11, 13\\%
33 &J1122+2124  &11 22 30.5&+21 24 45&3.46  &0.99      &2.91      &0.62       &0.69   &0.16 &~19.75         &11\\%
34 &J1146--5236 &11 46 30.1&-52 36 39&4.91  &0.99      &1.23      &--         &--     &--   &--             &6\\
35 &J1202+8522  &12 02 11.4&+85 22 18&3.95  &0.99      &0.61      &0.13       &0.69   &---  &--            &1, 4\\
36 &J1215--3905 &12 15 18.6&--39 05 17&10.5 &0.94      &1.43      &0.19       &0.90   &--   &--            &1, 6 \\
37 &J1236--3535 &12 36 45.1&--35 35 14&0.75 &0.99      &0.81      &0.09       &0.98   &0.07$^{b}$ &--            &1\\%
38 &J1240--3413 &12 40 03.7&--34 13 29&10.7 &0.99      &0.89      &--         &--     &0.07 &--            &1\\
39 &J1255--4447 &12 55 57.3&--44 47 54&5.12 &0.92      &0.53      &--         &--     &--   &--            &14\\
40 &J1259--4458 &12 59 45.6&--44 58 37&3.21 &0.99      &4.29      &--         &--     &--   &--            &7, 14\\
41 &J1303+3150  &13 03 15.5&+31 50 18&1.81  &0.61      &2.15      &0.20       &1.06   &0.16 &28.22          &1,2\\%
42 &J1306+4633  &13 06 45.7&+46 33 29&3.89  &0.99      &3.24      &0.21       &1.22   &0.22$^{b}$ &~48.89  &1, 2\\
43 &J1347--3905 &13 47 39.3&--39 05 42&7.23 &0.99      &0.42      &0.04       &1.05   &--   &--            &1\\
44 &J1409+7753  &14 09 26.0&+77 53 17&2.53  &0.99      &0.41      &0.11       &0.58   &---  &--            &1\\
45 &J1425+1210  &14 25 15.2&+12 10 09&4.47  &0.99      &1.38      &0.49       &0.46   &0.15 &~7.89         &11, 17\\
46 &J1629+0104  &16 29 27.3&+01 04 07&1.14  &0.99      &1.91      &0.31       &0.81   &---  &--            &1\\
47 &J1657--0148 &16 57 52.6&--01 48 01&5.88 &0.99      &1.97      &0.07       &1.49   &--   &--            &1 \\
48 &J1710+4239  &17 10 40.7&+42 39 45&6.94  &0.99      &1.54      &0.46       &0.54   &--   &--            &1\\%
49 &J1924--0032 &19 24 07.0&--00 32 07&6.94 &0.99      &1.05      &0.21       &0.72   &--   &--            &1, 8\\
50 &J1926+4831  &19 26 09.5&+48 31 28&3.13  &0.99      &1.19      &0.02       &1.82   &     &               &1, 4\\
51 &J1930--0312 &19 30 57.3&--03 12 56&8.96 &0.99      &0.80      &0.22       &0.57   &--   &--             &1,6 \\
52 &J2038--2011 &20 38 27.7&--20 11 07&1.13 &0.99      &2.98      &0.77       &0.60   &0.51 &--             &1 \\%
53 &J2058--2125 &20 58 03.8&--21 25 19&3.55 &0.97      &1.14      &0.11       &1.04   &--   &--             &1 \\

\hline
\end{tabular}

\end{threeparttable}
\end{table*}

\begin{table*}
\contcaption{\bf Candidate NAT radio sources}
\begin{threeparttable}
\begin{tabular}{ccccccccccccll}
\hline

Cat&  Name     &R.A-core  &Dec-core &Separation  &$\rho$    &$F_{150}$ &$F_{1400}$ & ${\alpha}^{1.4}_{0.15}$ & $z$ &$L_{150}$ & Other	\\
N0.&           &(J2000.0) &(J2000.0)&(arsec)     &       &(Jy)      &(Jy)       &($\pm0.05$)              &     & W$Hz^{-1}$& Catalog\\
   &           &          &         &            &       &          &           &                         &     &$\times 10^{25}$     \\
(1)& (2)       &(3)       &(4)      & (5)        &(6)    &(7)       &(8)        &(9)                     &(10) &(11)            &(12)  \\

\hline

54     &J2104+1916  &21 04 21.4 &+19 16 46&4.18 &0.99    &3.63      &0.42       &0.96   &--   &--             &1, 3\\
55     &J2137--4105 &21 37 51.4 &-41 05 19&7.31 &0.99    &2.19      &--         &--     &0.06 &--             &14 \\%
56     &J2144+2107  &21 44 06.4 &+21 07 44&2.87 &0.99    &0.38      &0.16       &0.38   &--   &--             &1, 8\\
57     &J2201--4427 &22 01 58.5&--44 27 06&2.33 &0.99    &1.11      &--         &--     &--   &--             &6, 14 \\
58     &J2227--3034 &22 27 54.5&--30 34 32&2.17 &0.99    &3.83      &0.89       & 0.65  &0.06 &~0.75          &1 \\
59     &J2319--1838 &23 19 30.0&--18 38 05&6.12 &0.99    &0.45      &0.15       &0.49   &--   &--             &1, 6 \\
60     &J2348--3117 &23 48 54.9&--31 17 32&8.80 &0.99    &1.14      &0.33       &0.55   &0.18 &~9.75          &1 \\
61     &J2351+0033  &23 51 54.4&+00 33 10&4.67  &0.95    &0.27      &0.08       &0.54   &0.27 &05.56          &1,11\\ %

\hline
\end{tabular}

\begin{tablenotes}
      \small
     \item 1: NVSS \citep{Con98}; 2: VLSS \citep{Co07}; 3: 4C \citep{Pi65, Go67, Ca69};  4: 6C \citep{Ba85, Ha88, Ha90, Ha91, Ha93a, Ha93b}; 5: 7C \citep{Mc90, Ko94, Wa96, Ve98}; 6: PMN \citep{Gri94}; 7: PKS \citep{Bo64}; 8: 87GB \citep{Gr91}; 9: B2 \citep{Co70, Co72, Co73, Fr74}; 10: B3 \citep{Fi85}; 11: VFK \citep{Va15}; 12: CRATES \citep{He07}; 13: FIRST \citep{Be95}; 14: SUMSS \citep{Mau03}:; 15: 2MASS \citep{Skr06}; 16: MG2\citep{Be86}; 17: MRC \citep{La81} ; 18: SSTSL \citep{Ra12}\\
$^{a}$ presents photometric redshift\\
$^{b}$ presents spectroscopic redshift\\  
    \end{tablenotes}
\end{threeparttable}
\end{table*}

\begin{table*}
\caption{Cluster details for candidate WAT radio source.}
	\label{tab:wat-cluster}
  \begin{threeparttable}
\begin{tabular}{cccccccccccccll}
\hline
	Cat &Name       &Cluster &$z_{cl}$ &($D_{c}$) &$\theta$ &D    &$m_{r}$ & $r_{500}$ &$R_{L}$ &$N_{500}$ &$M_{500}$&$\rho_{co}$ \\
	No  &           &        &         &(Mpc)     &($''$)   &(kpc)    &        &  (Mpc)     &        &   & $\times 10^{14}M_{\odot}$&(Mp$c^{-3}$) \\
\hline
1.  &J0003--3556&ABELL 2717   &0.05     &218.4     &036.9     &~133.8&--&--&--&--&--&-- \\
2.  &J0022+2317 &RM J002224.7+231733.0 &0.13      &556.6    &04.8 &~~41.2  &--&--&--&--&--&--     \\
3.  &J0030+1058 &NSCS J003033+105852   &--        &--       &78.0 &--      &--&--&--&--&-- &--\\      
4.  &J0054+3339 &ZwCl 0051.7+3322      &--        &--       &222.0&--      &--&--&--&--&--&--  \\
5.  &J0114+0029 &WHL J011425.6+002933  &0.34      &1377      &00.6  &~~10.7&18.36&0.82&39.20&19&2.25&53.0 \\
6.  &J0120+1451 &YSS2008 466           &0.05      &218.4    &162.0&~586.7  &--&--&--&--&--&--\\
7.  &J0123+3315 &MCXC J0123.6+3315     &0.02      &88.0     &021.6&~~31.6  &--&--&--&--&--&--       \\
8.  &J0128+3448 &WHL J012859.1+344842  &0.15      &639.0     &000.1 &~~~0.96&16.32&0.88&35.48&21&2.01 &58.6 \\
9.  &J0149+1403 &MSPM 01728            &0.07      &304.3    &003.6&~~17.8  &--&--&--&--&--&--       \\
10. &J0204+0415 &WHL J020431.3+041501  &0.14      &597.9    &038.4&~350.7  &16.2&0.60&14.9&9&0.87&25.1  \\
11.  &J0205--4124&SPT-CL J0205-4125     &--        &--       &258.0&--      &--&--&--&--&--&-- \\
12.  &J0209+0950 &NSC J020939+094834    &0.09      &389.3   &138.0 &~859.1  &--&--&--&--&--&--\\
13.  &J0257--0400&WHL J025743.7-035951   &0.18      &760.8     &010.8 &~121.3&18.9&0.80&31.8&19&2.00  &53.0 \\
14. &J0306--1206&ABELL 0415            &0.08      &346.9   &294.0 &1643.0  &--&--&--&--&--&-- \\
15. &J0315+0507 &WHL J031530.6+050746  &0.22      &920.2   &006.0 &~~78.9  &17.3&0.69&17.5&9&1.04 &25.1 \\
16. &J0317+1212 &WHL J031723.6+121234  &0.22      &920.2   &006.0 &~~78.9  &16.4&1.01&53.5&16&3.55 &44.7 \\
17. &J0522+2807 &WHL J052246.0+280640  &0.13      &556.6   &019.2 &~164.7  &14.7&0.89&39.6&16&2.55&44.7 \\
18. &J0631+2500 &PSZ2 G188.38+07.05    &0.08      &346.9   &055.8 &~311.8  &--  &--&--&--&--&-- \\
19. &J0708+7152 &WHL J070819.0+715224  &0.11      &473.4   &067.8 &~503.9  &15.3&--&57.2&--&3.49&-- \\
20. &J0733+4211 &WHL J073312.6+421156  &0.48      &1870.5  &006.0 &~132.2  &19.0&0.85&12.5&18&0.62&6.44 \\
21. &J0736+2412 &WHL J073618.1+241043  &0.14      &597.9   &107.0 &~977.4  &16.2&0.75&29.9&16&1.87 &44.7 \\
22. &J0757+3640 &WHL J075753.2+364022  &0.12      &515.1   &003.0 &~~24.1  &15.6&0.70&17.6&16&1.04 &44.7  \\
23. &J0805+1614 &WHL J080543.1+161356  &0.10      &431.4   &005.4 &~36.9   &15.8&0.76&21.7&14&1.31 &39.1 \\
24. &J0818+5437 &WHL J081803.9+543709  &0.10      &431.4   &018.0 &~123.0  &14.9&1.01&56.8&29&3.79  &81.0 \\
25. &J0856+4829 &WHL J085600.8+482910  &0.12      &515.1   &010.2 &~~81.8  &15.8&0.95&44.3&25&2.89  &69.8 \\
26  &J0912+1600 &NSC J091232+155800    &0.19      &801.0   &120.0 &1409.0  &--  &--  &--  &--&--&-- \\
27. &J0917+5509 &WHL J091708.0+550908  &0.19      &801.0     &000.2 &~~~2.4&16.27&0.79 &33.56 &15&1.89 &41.9 \\
28. &J0944+0247 &WHL J094443.2+024754    &0.21    &880.7     &000.4 &~~~5.07&16.59&1.12&91.99&46&5.62 &128.5 \\
29. &J1012+0841 &WHL J101206.4+084133    &0.09    &389.3     &000.2 &~~~1.3&15.78&0.65&11.72&10&0.52&27.93 \\
30. &J1015+1221 &WHL J101540.7+122030  &0.61      &2292.6  &031.2 &~774.8  &20.4&0.70&28.7&8 &1.79  &22.3\\
31. &J1019+7020 &WHL J101956.5+702034  &0.24      &--      &053.4 &--      &17.4&--  &70.1&--&4.37&-- \\
32. &J1022+5006 &WHL J102228.4+500620  &0.15      &639.0   &000.1 &~~~0.9&15.27&1.32&119.33&74&7.46 &206.7 \\
33. &J1032+3151 &WHL J103214.4+315214  &0.35      &1413.6  &054.6 &~996.9  &17.8&1.00&55.4&26&3.69&72.6 \\
34. &J1034+0736 &WHL J103409.1+073616  &0.45      &1768.2  &013.2 &~280.8  &--&--&--&--&--&-- \\
35. &J1042+0237 &NSC J104200+023739    &0.04      &175.2   &004.2 &~~12.3  &--&--&--&--&--&-- \\
36. &J1046--2911&ABELL S0646           &0.06      &261.5   &019.8 &~~85.0  &14.31&--&17.2 &--&0.93&-- \\
37. &J1048+3532 &RM J104851.5+353136.4 &0.36      &1450.0  &039.0 &~725.4  &19.1 &--&--   &--&--  &-- \\
38. &J1050+0432 &MaxBCG J162.734+4.525 &0.12      &515.1   &081.6 &~654.5  &--   &--&--   &--&--  &-- \\
39. &J1050--2405&MCXC J1050.6-2405     &0.20      &840.9   &026.4 &~322.4  &--   &--&--   &--&--  &-- \\
40. &J1051+1825 &NSC J105152+182302    &0.11      &473.4   &180.0 &1338.0  &--   &--&--   &--&--  &--\\
41. &1056+0255  &WHL J105617.0+025526  &0.39      &1558.0  &000.1 &~~~1.9&18.92&0.70&27.38&14&1.52 &39.1 \\
42. &J1058+0136 &SDSS-C4 1110          &0.04      &175.2   &053.4 &~156.7  &13.6 &--&28.9 &--&1.64&-- \\
43. &J1118+2754 &NSC J111853+275216    &0.06      &261.5   &132.0 &~566.9  &--   &--&     &  &--&-- \\
44. &J1119+6317 &WHL J111933.1+631717  &0.16      &679.8   &000.6 &~~~6.1&15.68&0.76&26.27&16&1.45 &44.7 \\
45. &J1120+2912 &WHL J112038.5+291234  &0.27      &1114.4  &002.2 &~~33.1  &17.0 &0.73&21.0&11&1.27 &30.7 \\
46. &J1130+2524 &WHL J113048.8+252436  &0.15      &639.0   &001.8 &~~17.4  &16.2 &0.71&20.5&16&1.23 &44.7 \\
47. &J1132+6311 &NSC J113242+631204    &0.10      &431.4   &060.6 &~414.2  &--   &--  &--  &--&--&-- \\
48. &J1142+1102 &WHL J114254.3+110133  &0.15      &639.0   &033.0 &~325.5  &16.0 &1.13&82.6&46&5.37&128.5 \\
49. &J1151+0422 &NSC J115145+042203    &0.13      &556.6   &000.4 &~~3.4&-- &-- &-- &-- &--&-- \\
50. &J1155+5755 &WHL J115556.3+575501  &0.16      &679.8   &029.4 &~300.2  &16.5 &0.84&28.0&18 &1.74&50.2 \\
51. &J1156+3432 &GMBCG J179.0225+34.55 &0.26      &1076.0  &029.4 &~437.9  &1.52 &--&--&--&--&-- \\
52. &J1202+5802 &WHL J120203.8+580207  &0.09      &389.3   &001.0 &~~~6.2  &15.2 &0.94&46.1&30&3.02 &83.8 \\
53. &J1205+3204 &WHL J120514.5+320417  &0.17      &720.4   &000.2 &~~~2.2&16.52&0.87&-- &-- &-- &-- \\
54. &J1206+3152 &WHL J120647.7+315231  &0.15      &639.0   &000.6 &~~~5.8&16.90&0.77&29.65&14&1.65 &39.1 \\
55. &J1242+5021 &WHL J124207.4+502147  &0.15      &639.0   &000.4 &~~~3.8&16.33&0.52&10.28&5&0.53 &13.96 \\
56. &J1249+0144 &WHL J124943.7+014447  &0.20      &840.9   &031.4 &~383.4  &16.9 &0.98&52.8&38&1.50 &106.1 \\
57. &J1304+6439 &WHL J130428.8+643937  &0.23      &959.4   &001.0 &~~13.6  &16.3 &0.79&29.3&13&1.83 &36.3 \\
58. &J1307+5651 &MaxBCGJ196.932+56.850 &0.23      &959.4   &000.3 &~~~4.1&-- &-- &-- &-- &--&-- \\
59. &J1311--0120&WHL J131132.1-011946  &0.17      &720.4   &015.0 &~161.0  &--&--&--&--&--&-- \\
60. &J1315+4841 &WHL J131527.6+484025  &0.48      &1870.5  &056.8 &1252.0  &19.8&1.03&81.9 &39&5.68 &108.9 \\
\hline
\end{tabular}
\end{threeparttable}
\end{table*}

\begin{table*}
    \contcaption{Cluster details for candidate WAT radio source.}
  \begin{threeparttable}
\begin{tabular}{cccccccccccccll}
\hline
	Cat &Name       &Cluster &$z_{cl}$ &($D_{c}$) &$\theta$ &D    &$m_{r}$ & $r_{500}$ &$R_{L}$ &$N_{500}$ &$M_{500}$&$\rho_{co}$ \\
	No  &           &        &         &(Mpc)     &($''$)   &(kpc)    &        &  (Mpc)     &        &   & $\times 10^{14}M_{\odot}$&(Mp$c^{-3}$) \\

\hline
61. &J1321+4235 &MSPM 05039            &0.08      &346.9   &083.5 &~468.0  &--&--&--&--&--&-- \\
62. &J1325+5736 &WHL J132511.2+573601  &0.12      &515.1   &006.6 &~52.9   &14.1 &1.03&109.4&59&7.81&164.8 \\
63. &J1331--0544&ABELL 1751            &0.15      &642.8   &090.0 &~243.7  &16.2 &--&34.2 &--&1.98 &-- \\
64. &J1412+7420 &ABELL 1893            &0.21      &880.7   &133.2 &1689.0  &--&--&--&--&--&-- \\
65. &J1414+0143 &MSPM 01989            &0.05      &218.4   &079.0 &~286.1  &--&--&--&--&--&--\\
66. &J1416+0219 &NSC J141606+021843    &0.15      &639.0   &108.1 &1046.8  &--&--&--&--&--&-- \\
67. &J1440+0328 &WBL 518               &0.02      &88.0    &012.1 &~~18.0  &--&--&--&--&--&-- \\
68. &J1449+3959 &WHL J144900.9+400044  &0.19      &801.0   &084.0 &~986.3  &16.8&0.66&18.5&13&1.10 &36.3 \\
69. &J1450+4418 &WHL J145039.8+441829  &0.27      &1114.4  &003.6 &~~55.1  &17.5&1.02&73.1&36&5.01 &100.5 \\
70. &J1509+3327 &WHL J150959.8+332746  &0.11      &473.4   &001.2 &~~10.1  &--&--&--&--&--&-- \\
71. &J1553+1530 &ZwCl 1551.4+1539      &0.14      &597.9   &001.0 &~~10.2  &--&--&--&--&--&-- \\
72. &J1604+2355 &WHL J160456.7+235558  &0.05      &218.4   &001.3 &~~~5.6  &12.7&--&15.0  &--&0.80&-- \\
73. &J1612+2929 &MSPM 00091               &0.03  &131.7     &083.0&~184.1&--&--&--&--&--&-- \\
74. &J1615+4711 &WHL J161541.3+471004     &0.20  &840.9     &104.6 &1278.8&17.0&0.83&33.3&20&2.11  &55.8  \\
75. &J1616+0926 &WHL J161653.2+092635    &0.20  &840.9     &000.8 &~~~9.8&-- &-- &-- &-- &--&-- \\
76. &J1636+2718 &WHL J163604.2+271829     &0.13  &556.6     &014.4 &~123.5&15.7&0.74&22.4&14&1.36  &39.1 \\
77. &J1711+1351 &WHL J171150.9+135151     &0.23  &959.4     &001.0 &~~13.6&16.1&0.83&33.6&12&2.13  &33.5  \\
78. &J1717+3734 &WHL J171725.4+373458  &0.09      &389.3     &000.6 &~~~3.7&-- &-- &-- &-- &-- &-- \\
79. &J1735+3137 &RM J173507.0+313755.5    &0.26  &1076.0    &007.5 &~130.3&17.8&--&26.8  &--&2.33 &-- \\
80. &J1930--1509&MCXC J1930.0-1509        &0.08  &346.9     &026.3 &~147.4&14.4&--&32.7  &--&1.88 &-- \\
81. &J2138+8307 &ABELL 2387               &0.14  &597.9     &029.3 &~268.1&--&--&--&--&--&-- \\
82. &J2139--3218&APMCC 693                &0.08  &348.9     &105.8 &~165.6&17.0&--&39.6  &--&2.33 &-- \\
83. &J2144--3105&EDCC 027                 &0.18  &765.4     &114.4 &~196.7&16.9&--&31.1  &--&1.78 &-- \\
84. &J2212+1304 &MaxBCG J333.226+13.076  &0.15  &639.0     &000.1 &~~~0.96&-- &-- &-- &-- &--&-- \\
85. &J2226+1721 &ABELL 2443               &0.11  &473.4     &062.6 &~465.3&14.9&--&52.5  &--&3.18 &-- \\
86. &J2300+1426 &WHL J230046.8+142602     &0.15  &639.0     &001.2 &~~10.6&--&--&--&--&--&-- \\
87. &J2310+0734 &Pegasus II CLUSTER       &0.04  &75.2      &058.0 &~~~12.0&13.3&--&45.8 &--&2.73 &-- \\
88. &J2322+4157 &ZwCl 2322.4+4157.6       &--    &--        &486.0 &--&--&--&--&--&--&--\\
89. &J2348+0043 &SDSS CE J357.08+00.73    &0.39  &1558.0    &061.0 &1192.7&--&--&--&--&--&-- \\

\hline
\end{tabular}
\end{threeparttable}
\end{table*}

\begin{table*}
    \caption{Cluster details for candidate NAT radio source.}
	\label{tab:nat-cluster}
  \begin{threeparttable}
\begin{tabular}{cccccccccccccll}
\hline
	Cat &Name       &Cluster &$z_{cl}$ &($D_{c}$) &$\theta$ &D    &$m_{r}$ & $r_{500}$ &$R_{L}$ &$N_{500}$ &$M_{500}$&$\rho_{co}$ \\
	No  &           &        &         &(Mpc)     &($''$)   &(kpc)    &        &  (Mpc)     &        &   & $\times 10^{14}M_{\odot}$&(Mp$c^{-3}$) \\
\hline

1.  &J0013--1930&WHY J001334.0-192902   &0.09      &389.3     &132.6 &~826.1&15.1&--&92.5&--&6.49  &-- \\
2.  &J0041--0922&ABELL 0085   &0.05     &218.4     &108.6     &~394.0&--&--&--&--&--&-- \\
3.  &J0041--4346&ABELL 2809   &0.15     &639.0     &190.8     &1849.4&15.9&--&23.0  &--&1.28 &-- \\
4.  &J0102--0050&NSCS J010240-005003    &0.25      &1037.3    &041.4 &~599.1&17.5&--&34.6&--&--&-- \\
5.  &J0148--3155&ABELL 2943   &0.14     &597.9     &016.8     &~153.4&16.3&--&88.3   &--&5.63  &-- \\
6.  &J0228--2814&ABELL 3023   &0.21     &880.7     &067.2     &~852.5&--&--&--&--&--&--\\
7.  &J0653+6919 &RX J0653.4+6919        &0.15      &639.0     &069.0 &~668.8&--&--&--&--&--&-- \\
8.  &J0704+6318 &ABELL 0556   &0.09     &389.3     &034.2     &~212.9&15.1&--&82.2 &--&5.20&-- \\
9.  &J0730+4051 &WHL J073045.4+405038   &0.12      &515.1     &060.0 &~481.2&--&--&--&--&--&-- \\
10. &J0735+2510 &WHL J073539.8+251020   &0.08      &346.9     &048.0 &~269.0&14.9&0.88&37.2&18&2.38 &50.2 \\
11. &J0802+6345 &NSC J080155+634523     &0.09      &389.3     &053.4  &~332.4&--&--&-- &--&--&-- \\
12. &J0953+7057 &ABELL 0875  &0.18      &760.8     &136.6     &1534.7&16.2&--&38.5&--&2.26&-- \\
13. &J1038--2453&PSZ2 G268.30+28.89     &0.12      &515.1     &054.2 &~434.7&15.4&--&--&--&  &-- \\
14. &J1111+4050 &ABELL 1190  &0.07      &304.3     &067.8     &~336.0&--&--&--&--&--&-- \\
15. &J1122+2124 &WHL J112229.9+212422   &0.15      &639.0     &022.8 &~221.0&16.4&1.16&83.7&45&5.82  &125.7 \\
16. &J1236--3535&ABELLs 0701  &0.07     &304.3     &196.2     &~972.4&14.4&--&33.2&--&1.92&-- \\
17. &J1240--3413&ABELL 3524   &0.07     &304.3     &020.4     &~101.1&14.0&--&32.7&--&1.89&-- \\
18. &J1303+3150 &ABELL 4056  &0.21      &880.7     &267.2     &--    &--  &--&-- &-- &--&--&--&--\\
19. &J1306+4633 &WHL J130650.0+463333   &0.24      &998.5     &044.4 &~623.3&16.1&1.41&165.7&87&12.33  &243.0 \\
20. &J1314+6220 &NSC J131425+621907     &0.13      &556.6     &036.6 &~314.1&--&--&--&--&--&-- \\
21. &J1409+7753 &NSC J140833+775227     &0.19      &801.0     &108.6 &1275.1&--&--&--&--&--&--\\
22. &J1425+1210 &NSC J142513+120946     &0.15      &639.0     &038.4 &~372.2&--&--&--&--&--&-- \\
23. &J1446--0846&ABELL 1964   &0.07     &304.3     &168.0     &~832.6&--&--&--&--&--&-- \\
24. &J1657--0148&RXSC J1657-0148        &0.03      &131.7     &040.8 &~~90.5&--&--&--&--&--&-- \\
25. &J1710+4239 &WHL J171040.7+423945   &0.17      &720.4     &014.4 &~154.5&16.2 &0.73&26.0&17&1.60 &47.5 \\
26. &J1926+4831 &CIZA J1926.1+4833      &0.09      &389.3     &067.2 &~418.3&15.3&--&58.0&--&3.54&-- \\
27. &J2022--2056&ABELLs 0868   &0.05    &218.4     &034.8     &~126.0&--&--&--&--&--&-- \\
28. &J2137--4105&APMCC 688     &0.06    &261.5     &252.0     &1082.4&--&--&--&--&--&-- \\
29. &J2227--3034&2PIGG J2227.0-3041     &0.07      &304.3      &002.4&~~11.8&--&--&--&--&--&-- \\
30. &J2348--3117&ABELL 4043  &0.18      &760.8     &044.0     &~494.3&16.4&--&66.5&--&4.12&-- \\ 
31  &J1930--0312&ABELL 3036  &--        &--        &329.0     &--    &--  &--&-- &-- &--&--&--&--\\

\hline
\end{tabular}
\end{threeparttable}
\end{table*}

\begin{table*}
    \caption{ Details of various cluster surveys used in the present work.The brief name is used to refer to the catalogues mentioned in this article. N is the number of clusters within the TGSS field without redshift,  $N_{z}$ is the number of those clusters with redshifts and $N_{t}$ is the total number of cluster.}
\label{tab:WAT-NAT-cat}
 \begin{threeparttable}
\begin{tabular}{cccccccll}
\hline
Sl  & Catalogue & Observation  &$N_{z}$   &$N$     &$N_t$&ref. \\
No  &  name    & band          &          &        &     &      \\
\hline
1.  &ABELL    &Optical &7      &--    &7         &\citet{Ab89}\\

2.  &RM       &Optical, IR &3 &-- &3  &\citet{Ro15}\\
                       
3.  &NSCS     & Optical  &2      &--    &2          &\citet{Lo04}\\
    
4.  &WHL      &Optical, X ray    &53  &--    &53         &\citet{We15}\\
   
5.  &ZwCl     &Optical   &1      &2    &3           &\citet{Zw61}\\

6.  &MCXC     & X ray    &3      &--    &3          &\citet{Pi11}\\

7.  &MSPM     &Optical   &4      &--    &4          &\citet{Sm12}\\
   
8.  &SPT-CL   &SZ        &--     &1    &1          &\citet{Bl15}\\
   
9.   &CIZA    & X ray    &1      &--    &1          &\citet{Eb02} \\
                                                           
10.  &NSC     &Optical   &8     &1     &9         &\citet{Gl09}\\

11.  &SDSS-C4 & Optical  &1      &--    &1          &\citet{Mi05}\\
     
12.  &MaxBCG  &Optical   &3      &--    &3         &\citet{Ko07}\\
  
13.  &GMBCG   &Optical   &1      &--    &1         &\citet{Ha10}  \\

14.  &APMCC   &Optical   &2     &--    &1          &\citet{Dal97}\\
     
15.  &EDCC    &Optical   &1    &--     &1          &\citet{Lu92} \\
                           
16.  &Pegasus &Optical   &1    &--     &1          &\citet{Ch76}  \\

                         
17.  &RX       &X ray    &1       &--   &1           &\citet{Mc04}\\
            
18.  &2PIGG    &IR       &1       &--  &1         &\citet{Eke04}  \\

19.  &YSS      &Optical  &1       &--  &1         &\citet{Yo08}  \\

20.  &PSZ2     & Radio   &2       &--   &2        &\citet{Kh16}  \\
     
21.  &RXSC     &X ray    &1       &--    &1       &\citet{Ch13}\\
                                  
22.  &WBL      &Optical &1        &--    &1       &\citet{Wh99}   \\
\hline
\end{tabular}
\end{threeparttable}
\end{table*}